# Impact of Engagement Allocation Across Social Platform Modalities on E-Commerce Performance

Gavin Wang
Purdue University

Yakov Bart
Northeastern University

Serguei Netessine
University of Pennsylvania

Lynn Wu
University of Pennsylvania

**Abstract:** Firms increasingly operate across multiple social media platforms, yet it remains unclear whether diversifying engagement across platforms enhances performance or simply fragments marketing efforts. We examine how the allocation of user engagement across platforms affects e-commerce sales performance. Using panel data on approximately 2,000 leading U.S. online retailers from 2012 to 2019, combined with detailed engagement measures across five major social media platforms, we construct firm-year indicators of engagement diversification and explore how they relate to sales performance. We find that greater diversification in engagement allocation is associated with significantly higher web sales. Importantly, this effect is not driven by platform adoption breadth or overall engagement volume. Rather than increasing traffic quantity, diversification improves conversion rates and enhances traffic quality. Mechanism analyses reveal that these performance gains stem from cross-modality complementarities: engagement distributed across heterogeneous content modalities (image, video, and mixed) generates reinforcing brand exposure, whereas diversification across platforms within the same modality yields limited benefits. Furthermore, the positive effects of diversification arise only when there is sufficient overlap in audiences across platforms, providing additional evidence for a memory-reinforcement mechanism. Taken together, these findings document the importance of engagement allocation structure and highlight the role of cross-modality complementarities in multi-platform digital marketing strategies.

# 1. Introduction

Beyond facilitating interpersonal interactions, social media platforms offer firms unprecedented opportunities to engage audiences through diverse content formats and interactive mechanisms (Aral et al., 2013; Bharadwaj et al., 2013; Rishika & Ramaprasad, 2019). With the proliferation of such platforms, firms increasingly maintain a presence across multiple platforms, devoting substantial resources to producing content, interacting with users, and building brand engagement.

Despite the ubiquity of multi-platform presence, it remains unclear how firms should allocate their engagement efforts across platforms. Prior research has largely examined firm activity on individual platforms, such as Facebook (e.g., Lee et al., 2018) or Twitter (Lambrecht et al., 2018; Wang et al., 2024), focusing on platform-specific marketing strategies. However, in practice, firms rarely operate in isolation on a single platform. Instead, they distribute engagement across multiple platforms that differ in user bases and content modalities — from image-centric environments such as Instagram and Pinterest to video-based environments such as YouTube and TikTok.

This multi-platform environment presents firms with a fundamental allocation challenge. On the one hand, concentrating engagement on a limited number of platforms may enhance focus and reduce coordination costs (Luo et al., 2020), and firms may benefit from specializing in platforms that best match their brand positioning (McIntyre, 2014). Digitally native brands such as Glossier initially relied heavily on Instagram to cultivate a strong online community[1], and firms like Shein have achieved rapid growth through intensive engagement on TikTok.[2] On the other hand, distributing engagement across platforms may generate complementary benefits: exposure to brand content across heterogeneous modalities — text, image, and video — may reinforce customer memory, enhance product evaluation, and improve purchase likelihood (Naik & Raman, 2003). Yet diversification may also fragment attention or dilute marketing resources (Sridhar & Sriram, 2015). Whether and how firms should diversify engagement across platforms is therefore a pressing managerial question.

Equally important is understanding *why* diversification might affect performance. It may shape sales through more than one pathway: expanding total traffic by broadening reach, or improving conversion efficiency by enhancing the quality and cognitive impact of brand exposure. Disentangling these mechanisms matters for both theory and practice.

In this paper, we investigate whether and why diversified social media engagement may improve e-commerce performance. Using a novel longitudinal dataset covering approximately 2,000 leading U.S. online retailers from 2012 to 2019, we construct detailed measures of engagement allocation across five major social platforms: Facebook, Twitter, Pinterest, Instagram, and YouTube. Our empirical strategy combines panel-data methods with an instrumental-variable approach to address endogeneity concerns, supported by a rich set of firm- and platform-level controls.

---

[1] See https://www.extole.com/blog/glossier-marketing-how-the-beauty-brand-used-word-of-mouth-to-shake-up-the-industry/.
[2] See https://www.vox.com/the-goods/22573682/shein-future-of-fast-fashion-explained.

We document three main findings. First, greater diversification in engagement allocation across platforms is associated with significantly higher web sales, and this effect is not driven by the breadth of platform adoption or total engagement volume. Second, diversification does not primarily increase traffic quantity; instead, it enhances conversion rates and improves traffic quality. Third, these gains reflect cross-modality complementarities: engagement distributed across heterogeneous content formats produces reinforcing effects consistent with a memory-reinforcement mechanism. Supporting this interpretation, the effect strengthens when audiences overlap across platforms — so that the same consumer is exposed to multiple modalities — and is concentrated among experience and hedonic products and among younger, higher-income customer segments. The results are robust across alternative measures and specifications.

Our study contributes to the information systems and digital marketing literature in several ways. First, we contribute to the literature on social media marketing, which has previously largely examined firm activity on individual platforms, by showing that *how* firms allocate engagement across platforms matters independently of how many platforms they use or how much they engage overall. The structure of allocation (not just its breadth or volume) shapes performance. Second, we extend work on multi-platform and multichannel marketing by identifying cross-modality complementarity as a distinct mechanism: gains arise from engagement distributed across heterogeneous content formats, not from expanding presence within the same modality. Third, we link this firm-level pattern to consumer-level outcomes, providing evidence that cross-modal exposure improves conversion efficiency and traffic quality rather than simply broadening reach.

The remainder of the paper proceeds as follows. Section 2 develops the theoretical framework and hypotheses. Section 3 describes the data and empirical methodology. Section 4 presents the main results. Section 5 concludes with implications and directions for future research.

## 2. Theory and Literature

### 2.1 Diversification of Multi-Platform Engagement and Sales Performance

Allocating marketing resources across communication channels has long been a central challenge for firms, and prior research suggests that the performance implications of such diversification depend on how channels interact with one another (Brynjolfsson et al., 2009; Goldfarb & Tucker, 2011; Li et al., 2018; Luo et al., 2020). One stream of work emphasizes substitution effects: investments in one medium may reduce the marginal effectiveness of investments in another, particularly when channels deliver inconsistent messages or when coordination across media becomes complex (Naik et al., 2005; Sridhar & Sriram, 2015; Sridhar et al., 2016). A second stream emphasizes complementarities: coordinated multi-channel exposure can reinforce brand awareness, increase information credibility, and enhance persuasion through repeated impressions across contexts (Naik & Raman, 2003; Chang & Thorson, 2004; Lim et al., 2015; Kumar et al., 2017). The mixed empirical findings in this literature reflect both mechanisms operating simultaneously, with the dominant effect depending on how informational cues from different channels interact (Danaher & Dagger, 2013; Sridhar et al., 2016).

Although prior research on social media has documented its effects on firm performance through brand visibility, peer influence, and electronic word-of-mouth (Li & Hitt, 2008; Kumar et al., 2016; Li & Wu, 2018; Bond et al., 2019; Wang et al., 2025), most studies examine activity within a single platform (e.g., Lee et al., 2018; Lambrecht et al., 2018) or aggregate social media into a single category alongside traditional channels (Danaher & Dagger, 2013; Kumar et al., 2017). In practice, however, consumers increasingly multihome across platforms (Tandoc Jr et al., 2019). By 2018, the average consumer had adopted at least eight social media accounts, nearly doubling the average from 2013.[3] This widespread multihoming leads to substantial overlap in platform usage and exposes users to similar content and trending information across platforms (de Meo et al., 2014). For example, it is reported that in 2022, over 70% of Twitter users also reported using Instagram.[4] Accordingly, firms maintain coordinated presences on multiple platforms simultaneously, each characterized by distinct user communities and content formats. Yet, despite this coordinated multichannel environment, a central question remains underexplored: how firms should optimally allocate engagement across platforms to maximize performance.

We argue that, holding total engagement constant, a more diversified allocation of engagement across platforms improves sales performance by increasing repeated exposure to related information across contexts. When consumers encounter similar brand content on multiple platforms, subsequent exposures reactivate prior impressions and strengthen memory encoding, thereby increasing the likelihood of favorable brand evaluations and purchase (Sridhar

[3] Source: https://www.statista.com/statistics/788084/number-of-social-media-accounts/.
[4] Source: https://x.com/stats_feed/status/1558862202806976514.

et al. 2016; Naik and Raman 2003). In social media environments, such reinforcement is particularly relevant because users commonly maintain accounts on multiple platforms, generating substantial audience overlap and repeated exposure opportunities across contexts. Accordingly, diversification increases the likelihood that a given consumer encounters related brand content multiple times across platforms. These reinforcement effects should be especially valuable in settings where consumers rely more heavily on cumulative and experiential information to evaluate products, such as for experience goods, whose quality is difficult to assess ex ante, and for consumers who depend more on digital signals when making purchase decisions. These arguments lead to our first hypothesis:

*H1: Holding the overall engagement volume constant, firms allocating engagement more diversely across social media platforms achieve higher sales performance than firms concentrating engagement on a smaller set of platforms.*

### 2.2 Cross-Modality Complementarity

While diversification across platforms may generate complementary exposure effects, the magnitude of these benefits depends on the heterogeneity of the platforms involved. Social media platforms differ not only in user composition but also in the dominant modalities through which information is communicated: Instagram and Pinterest emphasize images, YouTube emphasizes video, and Facebook and Twitter integrate multiple formats (De Vries et al., 2012; Voorveld et al., 2018).

Research in cognitive psychology suggests that information presented through multiple sensory modalities enhances memory formation and persuasion. When consumers process information through different perceptual channels, they develop richer cognitive representations of the underlying message (Paivio, 1986; Mayer & Moreno, 2003, Dijkstra et al., 2005). Visual imagery facilitates rapid recognition of product features, while video content provides richer demonstrations and narratives. For example, a consumer who first encounters an image-based post highlighting the aesthetic features of a new product on Instagram may later be exposed to a video-based campaign on YouTube that emphasizes brand narratives and emotional appeals (e.g., athlete storytelling in Nike's "Just Do It" campaigns). These multimodal cues reinforce one another, strengthening encoding and recall. By contrast, repeated exposure within a single modality tends to exhibit diminishing marginal effectiveness (advertising wear-out) because exposures activating similar cognitive pathways provide limited additional reinforcement (Campbell & Keller, 2003; Schmidt & Eisend, 2015). Empirical evidence from marketing supports this distinction: consumers exposed to multimedia campaigns exhibit higher credibility perceptions and stronger purchase intentions than those repeatedly exposed through a single medium (Chang & Thorson, 2004; Lim et al., 2015).

Applied to social media, this logic implies that the benefits of diversification depend on whether the platforms involved represent distinct modalities. A consumer who encounters an image-based post highlighting a product's aesthetic features on Instagram and later views a video-based campaign emphasizing brand narrative on YouTube receives complementary informational cues that engage distinct cognitive pathways, producing richer mental representations than either exposure alone. In contrast, diversification across platforms with similar modalities, such as engagement spread across Instagram and Pinterest, both image-centric, produces redundant cues. Although these platforms differ in user communities and interface, the informational signals they convey are similar, limiting the incremental value of additional exposures.

These arguments suggest that the effectiveness of diversification depends not only on the number of platforms used but also on the diversity of content modalities they represent:

*H2: Holding overall engagement volume constant, firms allocating engagement across platforms representing diverse content modalities achieve higher sales performance than firms diversifying engagement across platforms with similar modalities.*

## 3. Data and Methodology

The major dataset used in this study was obtained from the Internet Retailer Top 500 and Top 1000 Guide (Editions 2012 to 2019). The publisher of this guide, Vertical Web Media LLC., obtained comprehensive performance data from the top 1,000 U.S. e-commerce retailers each year[5], including total web sales, conversion rates, monthly unique visitors, multi-platform social media presence, and technology adoptions, etc., through a variety of channels, including direct surveying, collecting public information, and purchasing from third-party data providers, etc. The data report each year has been verified by multiple stakeholders to ensure authenticity, and all the retailers are given the opportunity to respond to the data before the reports are published. Data from this guide has been used previously in literature by Ayanso & Yoogalingam (2009) to study website functionalities and conversion rates and by Oberoi et al. (2017) to study the impact of technology outsourcing in retailing industries. Using eight years of this guide, we build an unbalanced panel data for about 7,000 observations, including about 2,000 unique companies since the top 1,000 list changes slightly every year.

[5] In 2012 and 2013 the number of surveyed companies was 500. From 2014 the number of surveyed companies has been 1,000.

The Internet Retailer Guide provides information on whether a retailer maintains an official account on major social media platforms (Facebook, Twitter, Pinterest, Instagram, and YouTube) as well as the total number of followers on each platform in each year. We complement this with more granular social media engagement data, including each firm's annual number of postings and likes on each platform[6], which was collected from the Wayback Machine (Internet Archive). More details about data extraction can be found in Appendix A.1. Through this process, we construct a novel firm-year panel dataset covering approximately 2,000 e-commerce retailers from 2012 onward, containing annual measures of followers, posting activity, and engagement outcomes (likes/pins) across five major social media platforms: Facebook, Twitter, Pinterest, Instagram, and YouTube.

### 3.1 Performance Measure

The main dependent variable in this study is the **Total Web Sales** (in U.S. dollars) of each retailer in each year. The total web sales metric is the total annual sales transacted online via desktop and mobile devices, provided directly by the Internet Retailer Guide. The total web sales metric has been widely used in literature to measure the performance of e-retailers (Ayanso & Mokaya, 2013; Ayanso & Lertwachara, 2015; Oberoi et al., 2017).

---

[6] Because engagement terminology differs across platforms (e.g., "likes" on Facebook and Instagram, "favorites" on Twitter in certain years, "saves" or "pins" on Pinterest), we use the term "likes" generically to denote platform-specific positive engagement outcomes.

To further unpack the mechanism through which diversified social media engagement affects firm performance, we decompose total web sales into three multiplicative components: **total website traffic** (i.e., the number of visits to the firm website), **conversion rate** (i.e., the percentage of visitors who complete a purchase), and **average ticket size** (i.e., the average dollar amount per transaction). By definition, total web sales equal the product of total traffic, conversion rate, and average ticket size. These three components are also reported in the Internet Retailer Guide and are constructed consistently across firms and years. This decomposition allows us to examine whether diversification primarily expands overall traffic, improves conversion efficiency, or increases transaction value.

### 3.2 Social Media Strategy

We obtain directly from the Internet Retailer Guide the presence data and, if present, the number of followers of each company's official handle each year on multiple major social platforms, including Facebook, Twitter, YouTube, Pinterest, and Instagram.[7] Based on the Pew Research Center's 2021 survey report, YouTube was the most popular social platform in the U.S., as 81% of U.S. adults have used it, followed by Facebook (69%), Instagram (40%) and Pinterest (31%)[8], so our dataset captures the most widely adopted social platforms in the sample period in the U.S. The presence data is a dummy variable where 0 stands for no presence and 1 for presence. Each year, we calculate the total number of present platforms that each e-retailer has (**Platform Number**) by adding up the presence variables.

---

[7] According to Internet Retailer Guide, the follower data was collected by a third-party data provider in September in each year and was verified by Vertical Web Media LLC.

[8] Source: https://www.pewresearch.org/internet/2021/04/07/social-media-use-in-2021/.

To measure the extent to which firms diversify their engagement across the 5 social media platforms, we construct **Engagement Diversification** as one minus the Herfindahl–Hirschman Index (HHI) of the distribution of likes received across platforms in a given year. This measure captures the dispersion of engagement allocation across platforms, rather than the depth of intensity on any single platform. The number of likes is used here because it directly measures users' voluntary reactions to platform content and is widely used in the social media literature as indicators of realized user engagement (De Vries et al., 2012; Lee et al., 2018; Wang et al., 2024). Specifically, $Engagement_Diversification_{i,t} = 1 - Likes_HHI_{i,t} = 1 - \sum_{p=1}^{5} likes_share_{i,t,p}^{2}$ where $likes_share_{i,t,p}$ stands for the share of the number of likes on the social platform $p$ for company $i$ in year $t$. For example, the share of the number of likes on Facebook for the company $i$ in year $t$ is: $likes_share_{i,t,p=Facebook} = \frac{\#\ Likes\ on\ Facebook_{i,t}}{\sum_{All\ platforms} \#\ Likes_{i,t}}$. To ensure comparability across platforms (e.g., the engagement intensity represented by 1 like on Instagram might be different from that on Twitter), we recenter the number of likes on each platform by dividing it by the platform-year median[9] before computing the shares. This adjustment reduces distortions arising from differences in platform size and engagement intensity. A higher diversification index suggests that the company's engagement is more diversified across multiple social platforms, while a lower diversification index suggests that the company is more concentrated on only a few platforms. For interpretability purposes, we standardize the diversification index by subtracting its sample mean and dividing by its sample standard deviation. This normalization allows the estimated coefficients to be interpreted as the effect of a one–standard deviation increase in diversification on firm performance.

[9] Median is used here to avoid skewness. The regression results remain substantially unchanged when metrics are recentered by dividing the platform-year means.

For robustness, we further construct 2 alternative indices to measure the extent to which firms diversify their engagement across platforms: (1) the **Audience Diversification**, which is calculated as one minus the HHI of the distribution of followers across social media platforms, and (2) the **Content Diversification**, which is computed as one minus the HHI of the annual number of postings published across platforms. We show in Appendix A.2 that the results are qualitatively unchanged across the three diversification indices. We further show in Appendix A.2 that the results remain consistent if substituting the HHI with the Gini Coefficient.

Beyond platform-level diversification, we further examine whether performance gains are driven by cross-modality complementarity (Voorveld et al., 2018; Naik & Raman, 2003). Consistent with prior literature documenting systematic differences in content formats and user interaction mechanisms across social media platforms, we categorize the five social media platforms into three modalities based on their dominant content format: Mixed Modality (Facebook and Twitter), Image-Centered Modality (Instagram and Pinterest), and Video-Centered Modality (YouTube) (De Vries et al., 2012). **Modality Diversification** is constructed as one minus the HHI of the distribution of likes received across these three modalities in a given year: $Modality_Diversification_{i,t} = 1 - ModalityHHI_{i,t} = 1 - \sum_{p=1}^{3} modality_share_{i,t,m}^{2}$ where $modality_share_{i,t,m}$ stands for the share of likes received within modality $m$ for company $i$ in year $t$. Similar to the construction of platform-level diversification, here we also recenter the platform-level likes to account for platform engagement differences and standardize the index by subtracting its sample mean and dividing by its sample standard deviation for interpretability purposes. Figure 1 plots the evolution of engagement diversification and modality diversification over time. Engagement diversification increased from 2012 to 2017 and declined modestly after 2018 (likely due to platform policy changes, such as Facebook's 2018 update), whereas modality diversification remained largely stable. A plausible explanation is that modality diversification reflects longer-term strategic choices tied to content production capabilities (e.g., image versus video), which are costly to adjust, whereas engagement allocation is more flexible and can be re-optimized in response to platform-level shocks.

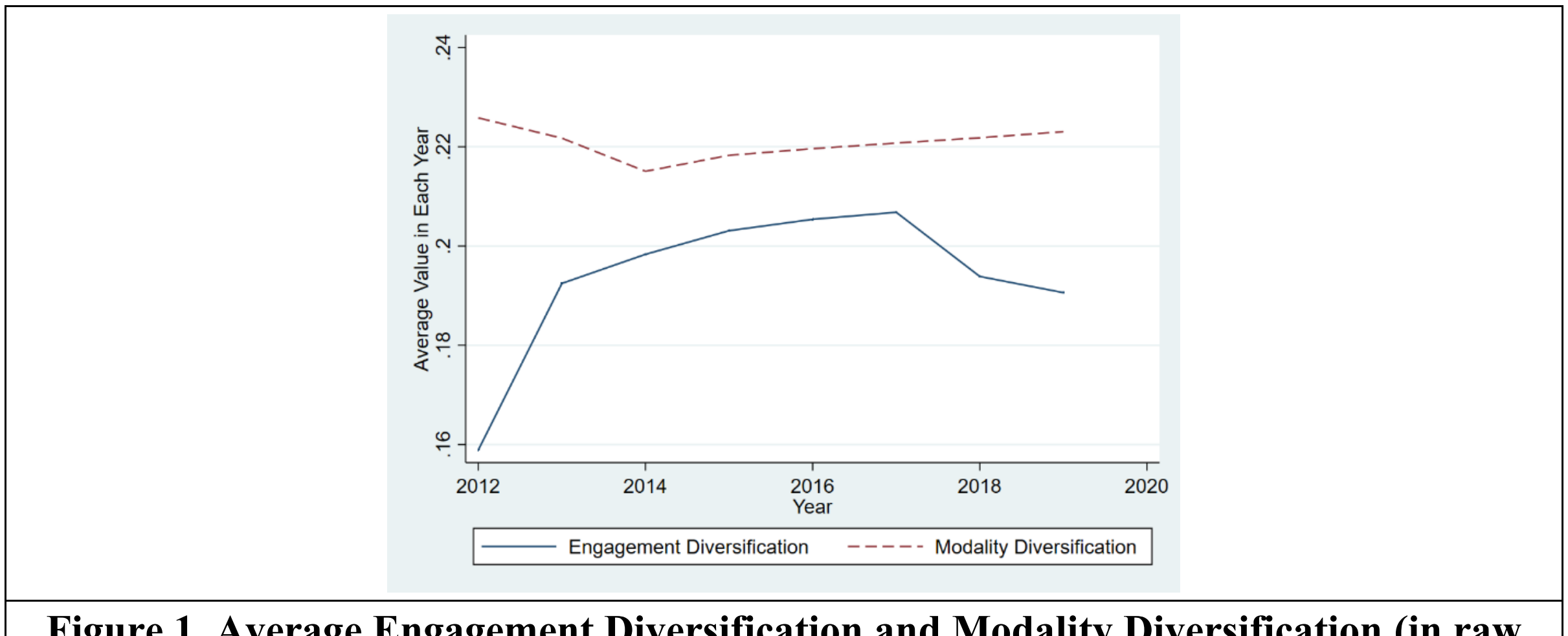

**Figure 1. Average Engagement Diversification and Modality Diversification (in raw values) in Each Year from 2012 to 2019**

In addition, to directly capture complementarities across platforms, we construct two pairwise interaction measures based on engagement shares. **Cross-Mode Complementarity** is defined as: $\sum_{(p,q)\in \text{CrossMode}} Share_{ipt} \times Share_{iqt}$, where $p, q$ are 2 different social media platforms belonging to different modalities, and $Share_{ipt}$ denotes the standardized share of likes received on platform $p$ by company $i$ in year $t$. **Same-Mode Complementarity** is constructed as $\sum_{(p,q)\in \text{SameMode}} Share_{ipt} \times Share_{iqt}$, where $p, q$ are 2 different social media platforms belonging to the same modality.

Conceptually, the cross-mode measure captures the extent to which a firm's engagement is jointly distributed across different content modalities (e.g., Instagram and YouTube, which belong to image and video modalities, separately), whereas the same-mode measure captures concentration within the same formats (e.g., Instagram and Pinterest, both belong to the image modality). If cognitive reinforcement operates across distinct sensory channels, we expect cross-mode complementarity to drive performance, whereas same-mode concentration should not generate similar benefits. For interpretability, these two metrics are also standardized by subtracting sample mean and dividing by sample standard deviation.

### 3.3 Other Variables

Both sales performance and social media engagement may be influenced by observable firm characteristics. We therefore include a rich set of control variables. First, firm performance may depend on merchant structure and product positioning (Lilien & Yoon, 1990; Oberoi et al., 2017). We control for dummy variables capturing **merchant type** (Catalog/Call Center, Consumer Brand Manufacturer, Retail Chain, Web Only) and **product category** in each firm-year. Importantly, the unit of observation in our dataset is the firm-year, and each firm-year is assigned one primary merchant type and one primary product category as reported by the Internet Retailer database. These classifications are provided directly by the data source rather than constructed by the authors, which helps ensure consistency and comparability across firms. Although firm fixed effects are included in all specifications, merchant type and product category are not fully absorbed because these characteristics can vary over time. For example, a retailer initially operating as a Web-Only firm may later expand into a retail chain format, and firms frequently broaden or shift their product mix as part of strategic repositioning. This pattern is common among digitally native brands (e.g., Warby Parker, Bonobos) that later expand into omnichannel retailing. Table 1 presents the distribution of merchant types and product categories in our sample.

**Table 1. Merchant Type and Product Categories (n=6,994)**

| Variable | Type or Category | No. of Observations | Percentage of Sample |
|---|---|---|---|
| Merchant Type | Catalog/Call Center | 888 | 12.70% |
| | Consumer Brand | 1,245 | 17.80% |
| | Retail Chain | 1,709 | 24.44% |
| | Web Only | 3,152 | 45.07% |
| Product Category | Apparel/Accessories | 1,792 | 25.62% |
| | Automotive | 223 | 3.19% |
| | Books/Music/Video | 254 | 3.63% |
| | Computers/Electronics | 512 | 7.32% |
| | Consumer Electronics | 96 | 1.37% |
| | Flowers/Gifts | 200 | 2.86% |
| | Food/Beverage | 37 | 0.53% |
| | Food/Drug | 289 | 4.13% |
| | Hardware/Home | 429 | 6.13% |
| | Health/Beauty | 375 | 5.36% |
| | Housewares/Home | 695 | 9.94% |
| | Jewelry | 251 | 3.59% |
| | Mass Merchant | 397 | 5.68% |
| | Office Supplies | 279 | 3.99% |
| | Specialty | 392 | 5.60% |
| | Specialty/Non-Apparel | 46 | 0.66% |
| | Sporting Goods | 482 | 6.89% |
| | Toys/Hobbies | 245 | 3.50% |

Beyond serving as controls, product characteristics also provide a natural setting for examining the heterogeneous impact of diversification. If diversification enhances consumer evaluation through richer sensory cues and affective processing, its performance impact should vary systematically across product types.

Second, we control for brand popularity and scale. Specifically, we control for firm age (**years since launch**) and firm size measured by the number of stock-keeping units available online (**SKUs**), as larger and more established firms may have greater operational capacity and marketing resources.

Third, to ensure that the estimated diversification effects measure allocation effects and are not driven by overall social media scale or specific platform intensity, we include several controls capturing aggregate social media presence and activity. These include: (1) the total platform number on which the retailer maintains an official account, (2) the total number of followers across all platforms, (3) the total number of postings across platforms, (4) the total number of engagements (likes or equivalent actions) across platforms, and (5) the share of engagement on each of the five individual platforms. These controls ensure that our diversification measures capture how engagement is distributed across platforms, rather than reflecting the influence of any single platform or the overall magnitude of social media activity.

### 3.4 Summary Statistics

Table 2 reports the summary statistics for the data. Our data includes 6,994 observations for about 2,000 unique online retailers over eight years.[10]

**Table 2. Summary Statistics for Internet Retailer Data (n=6,994)**

| Statistic | Mean | St. Dev. | Min | Max |
|---|---|---|---|---|
| Log Web Sales (dollars) | 17.88 | 1.57 | 12.49 | 25.92 |
| Log Conversion Rate | -3.71 | 0.70 | -9.21 | -0.99 |
| Log Total Traffic | 15.51 | 2.10 | 7.18 | 24.11 |

[10] Due to space constraints, the correlation matrix is available in Appendix A.5.

| Log Ticket Size (dollars) | 4.96 | 0.87 | 0.69 | 9.22 |
| --- | --- | --- | --- | --- |
| Engagement Diversification | 0 | 1 | -2.17 | 3.51 |
| Audience Diversification | 0 | 1 | -1.34 | 2.36 |
| Content Diversification | 0 | 1 | -3.34 | 2.40 |
| Modality Diversification | 0 | 1 | -1.98 | 3.28 |
| Cross-Mode Complementarity | 0 | 1 | -3.59 | 3.30 |
| Same-Mode Complementarity | 0 | 1 | -1.96 | 2.72 |
| has Facebook | 0.98 | 0.13 | 0 | 1 |
| has Twitter | 0.96 | 0.19 | 0 | 1 |
| has YouTube | 0.87 | 0.34 | 0 | 1 |
| has Pinterest | 0.76 | 0.43 | 0 | 1 |
| has Instagram | 0.57 | 0.50 | 0 | 1 |
| Facebook Engagement Share | 0.17 | 0.12 | 0 | 0.84 |
| Twitter Engagement Share | 0.25 | 0.09 | 0 | 0.85 |
| YouTube Engagement Share | 0.07 | 0.10 | 0 | 0.99 |
| Pinterest Engagement Share | 0.31 | 0.18 | 0 | 0.83 |
| Instagram Engagement Share | 0.21 | 0.10 | 0 | 0.88 |
| Platform Number | 4.14 | 1.06 | 0 | 5 |
| Log Total Number of Followers | 13.18 | 3.38 | 0 | 24.10 |
| Log Total Number of Postings | 7.16 | 0.38 | 5.75 | 9.32 |
| Log Total Number of Engagement | 13.42 | 0.93 | 12.72 | 17.61 |
| Age | 12.53 | 6.50 | 1 | 42 |
| SKUs on Web | 3.5e6 | 3.7e7 | 10 | 1e9 |

## 3.5 Empirical Methodology

### *3.5.1 Baseline Regression Models*

To examine whether diversified engagement allocation across social media platforms improves e-commerce performance, we estimate the following baseline model with retailer fixed effects ($\gamma_i$) and time dummies ($\tau_t$):

$$\log WebSales_{i,t} = \beta_1 Engagement_Diversification_{i,t} + Controls + \gamma_i + \tau_t + \varepsilon_{i,t}$$

The distribution of total web sales is positively skewed; consistent with prior studies (e.g., Duan et al., 2008; Oberoi et al., 2017), we apply a natural logarithmic transformation to normalize the dependent variable and mitigate the influence of extreme values.[11]

To formally examine the performance benefits of diversified social media modalities, we

[11] The sample skewness of Web Sales is 31.5. After taking the natural logarithm, the sample skewness decreased to 0.67.

estimate two sets of interaction models below:

***(1) Modality-Level Reinforcement***

First, we test whether the effect of overall social media diversification is strengthened when engagement is more diversely distributed across distinct content modalities:

$$\begin{aligned}\log WebSales_{i,t} = {} & \beta_1 Engagement_Diversification_{i,t} + \beta_2 Modality_Diversification_{i,t} \\ & + \beta_3 Engagement_Diversification_{i,t} \times Modality_Diversification_{i,t} + Controls + \gamma_i + \tau_t \\ & + \varepsilon_{i,t}\end{aligned}$$

In this specification, Modality Diversification captures the extent to which engagement is diversified across different content formats (mixed, image, and video). The interaction term allows us to test whether platform-level diversification generates stronger performance gains when accompanied by cross-modality diversification. A positive and significant $\beta_3$ would suggest that engagement diversification is particularly effective when it spans heterogeneous content modalities, consistent with a cognitive reinforcement mechanism.

***(2) Cross-Mode versus Same-Mode Complementarity***

To further disentangle the source of complementarities, we replace modality diversification with pairwise complementarity measures:

$$\begin{aligned}\log WebSales_{i,t} = {} & \beta_1 Engagement_Diversification_{i,t} + \beta_2 Cross_Mode_Complementarity_{i,t} \\ & + \beta_3 Engagement_Diversification_{i,t} \times Cross_Mode_Complementarity_{i,t} \\ & + \beta_4 Same_Mode_Complementarity_{i,t} \\ & + \beta_5 Engagement_Diversification_{i,t} \times Same_Mode_Complementarity_{i,t} + Controls + \gamma_i \\ & + \tau_t + \varepsilon_{i,t}\end{aligned}$$

Here, Cross-Mode Complementarity measures the joint distribution of engagement shares across platforms belonging to different modalities, whereas Same-Mode Complementarity captures joint distribution within the same modality. This specification allows us to directly test whether diversification generates value through heterogeneity in content formats rather than through concentration within similar formats. A positive and significant $\beta_2$ and $\beta_3$ would suggest that engagement diversification is more beneficial when it spans distinct modalities. No positive effect (or a weaker effect) for $\beta_4$ and $\beta_5$ would provide stronger evidence that performance gains arise from cross-modality complementarity rather than general engagement dispersion.[12]

### *3.5.2 Identification Strategies*

Due to the endogeneity of organizational practices in the archival data, providing conclusive causal evidence is challenging. In our research context, endogeneity may arise because firms adjust their social media strategies in response to prior marketing performance and anticipated demand conditions, as well as from time-varying unobserved managerial preferences. Where possible, we use two-way fixed effects regression models to control for both firm-level time-invariant heterogeneity and common time trends, and include a rich set of controls to account for observable factors. However, these approaches may not fully address remaining concerns related to reverse causality and time-varying unobservables. To mitigate these issues, we further adopt instrumental variable strategies to isolate exogenous variation in firms' social media strategies.

#### *(1) Platform Policy Change as Identification*

The first set of instrumental variables exploits policy shocks on major social media platforms that plausibly alter firms' incentives to concentrate or diversify their social media engagement across platforms. Specifically, we leverage several platform-level policy changes during our sample period that affect the relative attractiveness of particular platforms. These changes were implemented unilaterally by the platforms and applied simultaneously to all users, often with little advance notice, making them plausibly exogenous to individual firms' social media strategies and marketing decisions. For example, Facebook's 2018 News Feed algorithm update prioritized "meaningful interactions from close connections" among users and reduced the reach of business content, thereby weakening firms' incentives to concentrate engagement on Facebook.[13] In contrast, Pinterest's introduction of Buyable Pins in 2015 enhanced the platform's e-commerce functionality and increased incentives for retailers to concentrate activity on Pinterest.[14] Following a standard exposure–shock design (Goldsmith-Pinkham et al., 2020), we interact these policy changes with firms' pre-existing platform exposure, measured by their engagement share on the affected platform prior to the policy change. This is because the platform policy changes vary only over time and are common to all firms, while firms differ in their pre-existing exposure to the affected platforms. As a result, firms with higher pre-existing engagement on a given platform are more strongly affected by the corresponding policy change. The first-stage F-statistics for both policy changes are all above 20, suggesting that these policies meaningfully changed retailers' strategies of engagement diversification.

[12] All variables entering interaction terms are mean-centered prior to estimation to mitigate multicollinearity. The same applies below.

We also conduct placebo tests using platform changes that are unlikely to affect firms' engagement allocation. For instance, YouTube's introduction of YouTube Red (an ad-free viewing service) in 2015 primarily altered user viewing experiences without materially affecting firms' content reach or marketing incentives.[15] Consistent with this interpretation, exposure to this policy does not significantly affect engagement diversification. These null results help rule out confounding time trends or coincidental platform changes and support the validity of our identification strategy.

[13] See: https://about.fb.com/news/2018/01/news-feed-fyi-bringing-people-closer-together/.
[14] See: https://www.theverge.com/2015/6/2/8710309/pinterest-buy-buttons.

These platform policy changes also generate exogenous variation in modality diversification. Because each policy targets a specific platform, and platforms differ in their dominant content modality, shifts in platform engagement translate into changes in the allocation of engagement across modalities. We exploit this variation to identify the role of modality diversification and cross-modality complementarity in firms' social media strategies.

To further assess instrument relevance, we examine heterogeneity in the first-stage relationship by firms' pre-existing platform exposure. We find that the instruments significantly predict engagement diversification only among firms with higher prior exposure to the affected platform, while the relationship is weak and insignificant among low-exposure firms (see Appendix A.3). This pattern suggests that the instruments are most relevant precisely where the underlying shocks are economically meaningful, and, together with the null effect for the YouTube Red policy, helps rule out confounding time trends or general platform dynamics.

With respect to the exclusion restriction, our identification relies on the fact that these policy changes were externally imposed and largely unanticipated. In addition, we control for platform-specific engagement levels and overall social media activity, so the identifying variation comes from how firms reallocate engagement across platforms rather than from the overall intensity of engagement.

[15] See: https://blog.youtube/news-and-events/red/.

***(2) Hausman-type Instruments***

To strengthen identification and avoid reliance on a single source of variation, we further construct Hausman-type instruments for our key endogenous variables, including engagement diversification, modality diversification, and cross-mode complementarity, based on geographically proximate firms' social media strategies. Specifically, for each firm $i$ in state $s$ and year $t$, we compute (i) the average engagement diversification, (ii) the average modality diversification, and (iii) the average cross-mode complementarity of other e-retailers located in the same state but operating in different product categories (Hausman, 1996; Nevo, 2001). Each of these state-level averages serves as an instrument for the corresponding firm-level variable. Figure 2 shows the geographical distribution of the retailers in the sample.

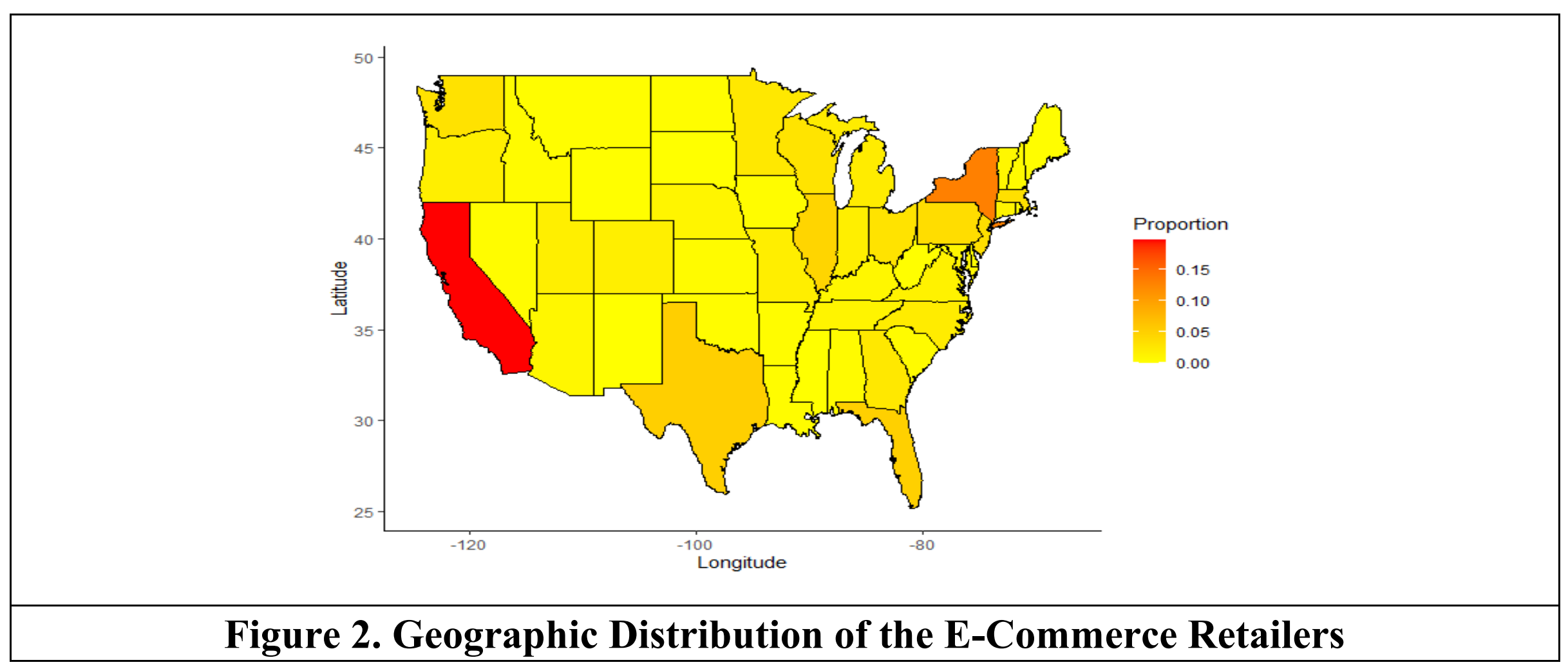

**Figure 2. Geographic Distribution of the E-Commerce Retailers**

This approach leverages the fact that the diffusion of digital practices and social media strategies is correlated within geographic regions due to shared infrastructure, labor markets, and local knowledge spillovers (Forman et al., 2002, 2008; Kulshrestha et al., 2012; Kim et al., 2013), while firms in different product categories are unlikely to directly affect each other's sales outcomes. Accordingly, identification arises from exposure to local peers' strategies while excluding within-category competitors. Our results are also robust to excluding firms in large states such as California and New York (see Appendix A.4), alleviating concerns that the findings are driven by high-performing regions.

For both the platform policy changes and the Hausman-type instruments, the first-stage F-statistics exceed 20 in all 2SLS specifications, mitigating concerns about weak instruments. While no instrumental variable strategy can fully eliminate all concerns about endogeneity, the consistency of results across multiple identification strategies provides suggestive causal evidence of our findings. Thus, we interpret the evidence as identifying economically meaningful and plausibly causal variation in social media diversification strategies, while recognizing the remaining limitations inherent in observational firm-level data.

## 4. Results and Interpretation

### 4.1 Diversified Engagement Is Associated with Higher Sales

We first show model-free evidence that engagement diversification is associated with higher sales performance. Figure 3 compares the average web sales with 95% confidence intervals of firms in the top and bottom diversification quintiles over time. Firms in the top diversification quintile consistently exhibit higher average sales than those in the bottom quintile.

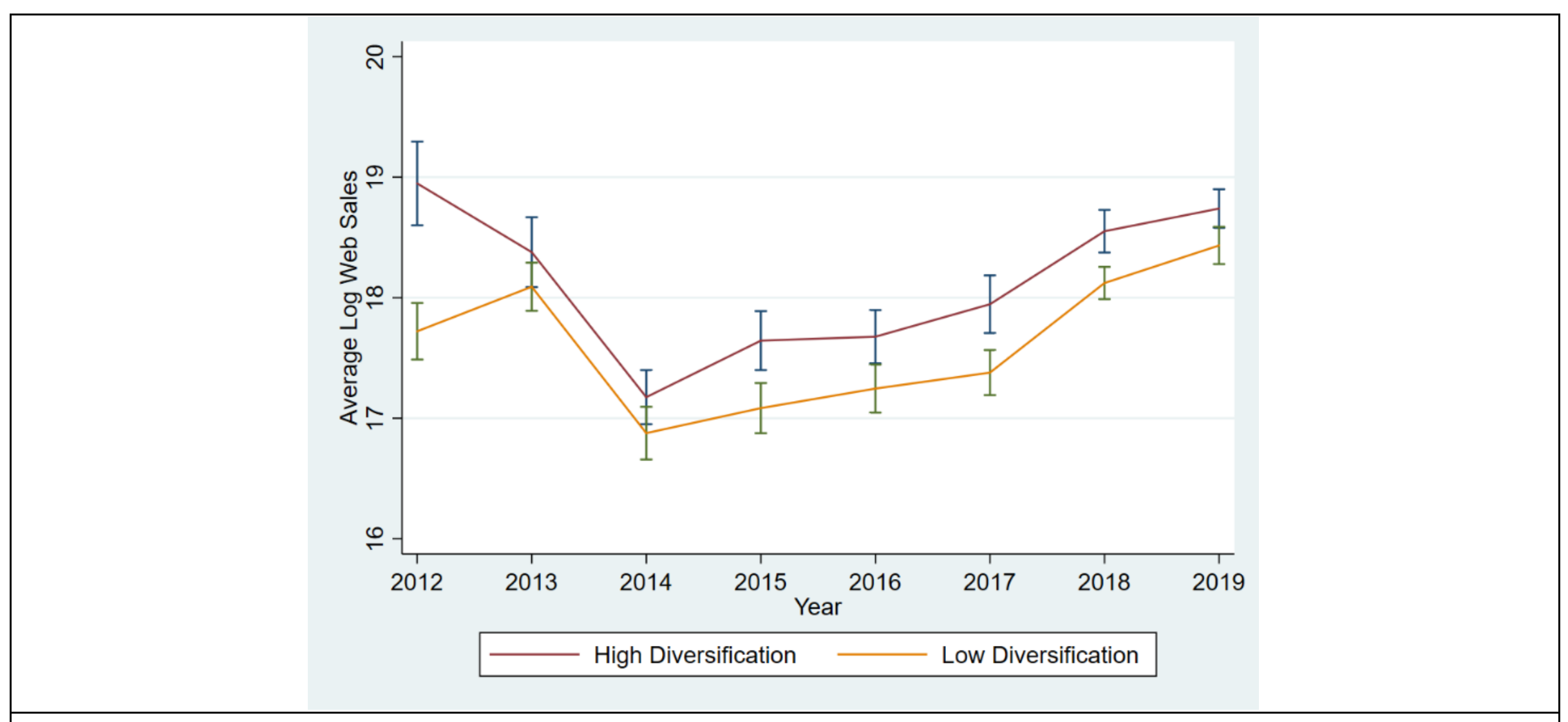

**Figure 3. Average Log Web Sales with 95% Confidence Intervals of Companies in High vs. Low Engagement Diversification Levels in Each Year from 2012 to 2019**

We examine the effect of engagement diversification on e-retailers' web sales in Table 3. Column (1) reports OLS estimates controlling for overall social media activity (total platforms, followers, posts, and likes) and firm characteristics, while Column (2) further controls for platform-specific engagement shares. Columns (3) and (4) present 2SLS estimates using platform policy shocks and Hausman-type instruments, respectively. Across all specifications, engagement diversification is positively associated with sales performance. A one-standard-deviation increase in diversification is associated with a 3.5%–4.3% increase in web sales in the OLS models, and the effect remains statistically significant under 2SLS.

The IV estimates differ modestly across instruments, reflecting variation in the populations affected. Policy shocks primarily impact firms with greater platform dependence and thus yield larger effects, whereas Hausman-type instruments capture more gradual, peer-driven adoption and produce smaller estimates. Nevertheless, the estimated coefficients are statistically indistinguishable based on t-tests.

Across all columns, the platform number is insignificant, indicating that the benefits of diversification are not driven by platform breadth per se, but by how engagement is allocated. Platform-specific engagement shares exhibit heterogeneous effects, yet diversification remains positive and significant after controlling for these shares, suggesting that the results are not driven by any single platform. Variance inflation factor diagnostics (mean VIF = 2.67) indicate that multicollinearity is not a concern.

Taken together, these results highlight the importance of the allocation structure of social media engagement, providing support for H1. The findings are robust to matched-sample estimations and to excluding firms in large states such as New York and California (see Appendix A.2 and A.4).

**Table 3. Social Media Engagement Diversification Effect on Web Sales**

| Model | OLS | OLS | Policy IV 2SLS | Hausman IV 2SLS |
|---|---|---|---|---|
| DV: Log web sales | (1) | (2) | (3) | (4) |
| | | | | |
| Engagement Diversification | 0.043*** | 0.035*** | **0.042**** | **0.032***** |
| | (0.010) | (0.010) | **(0.021)** | **(0.011)** |
| Share of Facebook Engagement | | 0.079 | 0.089 | 0.075 |
| | | (0.117) | (0.119) | (0.117) |
| Share of Twitter Engagement | | -0.218** | -0.216** | -0.219** |
| | | (0.101) | (0.101) | (0.101) |
| Share of Instagram Engagement | | 0.347*** | 0.337*** | 0.351*** |
| | | (0.082) | (0.085) | (0.082) |
| Share of YouTube Engagement | | -0.159* | -0.153* | -0.161* |
| | | (0.090) | (0.091) | (0.090) |
| Platform Number | 0.014 | 0.014 | 0.012 | 0.015 |
| | (0.012) | (0.012) | (0.013) | (0.012) |
| Log Total Number of Followers | 0.015** | 0.018*** | 0.021** | 0.017** |
| | (0.006) | (0.006) | (0.009) | (0.007) |

| Log Total Number of Postings | 0.082*** | 0.077*** | 0.074** | 0.079*** |
|---|---|---|---|---|
| | (0.029) | (0.029) | (0.031) | (0.029) |
| Log Total Number of Engagements | 0.210*** | 0.193*** | 0.190*** | 0.194*** |
| | (0.021) | (0.027) | (0.028) | (0.027) |
| Other Controls | Company Age, Number of SKUs, Retailer FE, Year FE, State-Year FE, Merchant Type, Product Category | | | |
| Observations | 6,994 | 6,994 | 6,994 | 6,994 |
| R Squared | 0.970 | 0.971 | \ | \ |

Notes:

1. The dependent variable is the log total web sales of the focal firm in the focal year.
2. The Engagement Diversification is constructed as one minus the HHI of the distribution of likes received across platforms in a given year. Higher values indicate that user engagement is broadly distributed rather than concentrated on a single platform.
3. Share of Facebook/Twitter/Instagram/YouTube Engagement is calculated as the ratio of the number of likes received on a single platform over the total number of likes received from all platforms in a given year. To ensure comparability across platforms, we standardize each metric by dividing it by the platform-year median before computing the share ratio. This adjustment reduces distortions arising from differences in platform size and engagement intensity. The Share of Pinterest Engagement has dropped because of multicollinearity (all shares add up to 1).
4. Platform Number measures how many social platforms (among Facebook, Twitter, Instagram, YouTube, Pinterest, value range 0~5) that the focal retailer has adopted in the given year.
5. Columns (1) and (2) report the OLS results with company and year fixed effects. Column (3) reports the 2SLS results using platform policies as instrumental variables. Column (4) reports the 2SLS results using Hausman-type instruments as instrumental variables. The first-stage F-statistic is greater than 20. Variables in bold are treated as endogenous and estimated using IV 2SLS.
6. Clustered standard errors on firm level in parentheses: *** p<0.01, ** p<0.05, * p<0.1.

## 4.2 Cross-Modality Complementarity

We next examine the role of modality heterogeneity in shaping the performance impact of diversified social media engagement. Table 4 reports the results. Across all specifications, we control for platform-level engagement shares, overall platform adoption, total social media activity, and firm characteristics. Column (1) shows that both engagement diversification and modality diversification are positively associated with performance. Column (2) introduces their interaction, which is significantly positive, indicating that diversification is more valuable when engagement is distributed across heterogeneous content modalities. The main effect of engagement diversification becomes negative and insignificant, suggesting that its benefits primarily arise from modality heterogeneity rather than diversification per se. These results remain directionally consistent under the 2SLS specifications in Columns (3) and (4). The sign flip of engagement diversification after introducing modality diversification suggests that diversification confined to platforms with similar modalities may be ineffective or even costly, likely due to redundant content provision. In Column (5), we replace modality diversification with a measure of cross-mode complementarity. The interaction between engagement diversification and cross-mode complementarity is positive and significant, while the main effect of diversification remains insignificant. Columns (6) and (7) confirm these findings under the instrumental variable approach. Finally, Columns (8) and (9) further include the same-mode complementarity. Column (9) shows that cross-mode complementarity is positively associated with performance, whereas same-mode complementarity enters with a negative coefficient. The two measures are not strongly correlated, indicating that they capture distinct dimensions of engagement structure. Multicollinearity is not a concern (mean VIF = 3.02).

Taken together, these results suggest that not all forms of diversification are equally beneficial: cross-modality combinations generate complementarities, whereas concentration within similar modalities may lead to limited returns due to redundant information. These findings provide support for H2.

**Table 4. Cross-Modality Complementarities on Social Media Platforms**

| Model | OLS | OLS | Policy IV 2SLS | Hausman IV 2SLS | OLS | Policy IV 2SLS | Hausman IV 2SLS | OLS | OLS |
|---|---|---|---|---|---|---|---|---|---|
| DV: Log web sales | (1) | (2) | (3) | (4) | (5) | (6) | (7) | (8) | (9) |
| | | | | | | | | | |
| Engagement Diversification | 0.021** | -0.006 | **-0.152** | **-0.030*** | 0.004 | **-0.175** | **-0.028** | 0.034*** | -0.020 |
| | (0.011) | (0.012) | **(0.094)** | **(0.018)** | (0.012) | **(0.108)** | **(0.017)** | (0.010) | (0.013) |
| Modality Diversification | 0.027*** | 0.051*** | **0.178**** | **0.071***** | | | | | |
| | (0.008) | (0.010) | **(0.082)** | **(0.015)** | | | | | |
| Modality Diversification * Engagement Diversification | | 0.022*** | **0.139**** | **0.041***** | | | | | |
| | | (0.005) | **(0.070)** | **(0.011)** | | | | | |
| Cross-Mode Complementarity | | | | | 0.038*** | **0.187**** | **0.065***** | | 0.057*** |
| | | | | | (0.009) | **(0.090)** | **(0.013)** | | (0.009) |
| Cross-Mode Complementarity * Engagement Diversification | | | | | 0.017*** | **0.170**** | **0.045***** | | 0.021*** |
| | | | | | (0.005) | **(0.084)** | **(0.012)** | | (0.005) |
| Same-Mode Complementarity | | | | | | | | -0.013 | -0.028*** |
| | | | | | | | | (0.008) | (0.007) |
| Same-Mode Complementarity * Engagement Diversification | | | | | | | | -0.001 | -0.009*** |
| | | | | | | | | (0.006) | (0.002) |
| Other Controls | Share of Facebook/Twitter/Instagram/YouTube Engagement, Platform Number, Log Total Number of Followers, Log Total Number of Postings, Log Total Number of Engagements, Company Age, Number of SKUs, Retailer FE, Year FE, State-Year FE, Merchant Type, Product Category, Product Category-Year FE | | | | | | | | |
| Observations | 6,994 | 6,994 | 6,994 | 6,994 | 6,994 | 6,994 | 6,994 | 6,994 | 6,994 |
| R Squared | 0.970 | 0.971 | \ | \ | 0.971 | \ | \ | 0.970 | 0.971 |

Notes:

1. The dependent variable is the log total web sales of the focal firm in the focal year.
2. Engagement Diversification is constructed using the same method as in Table 3.
3. Modality Diversification is constructed as one minus the HHI of the distribution of likes received across 3 modalities (Mixed Modality, including Facebook and Twitter; Image Modality, including Instagram and Pinterest; Video Modality, including YouTube) in a given year. To ensure comparability across platforms, we standardize the distribution of likes on each platform by dividing it by the platform-year median before computing the HHI. This adjustment reduces distortions arising from differences in platform size and engagement intensity. The same standardization is also applied below when calculating cross-mode complementarity and same-mode complementarity.
4. Cross-Mode Complementarity is constructed as $\sum_{(p,q)\in \text{CrossMode}} Share_{ipt} \times Share_{iqt}$, where $p, q$ are 2 different social media platforms belonging to different modalities, and $Share_{ipt}$ represents the share of likes received on the platform $p$ by company $i$ in year $t$. Same-Mode Complementarity is constructed as $\sum_{(p,q)\in \text{SameMode}} Share_{ipt} \times Share_{iqt}$, where $p, q$ are 2 different social media platforms belonging to the same modality.
5. Columns (1), (2), (5), (8), and (9) report the OLS results with company and year fixed effects. Columns (3) and (6) report the 2SLS results using platform policies as instrumental variables. Columns (4) and (7) report the 2SLS results using Hausman-type instruments as instrumental variables. The first-stage F-statistic is greater than 20. Variables in bold are treated as endogenous and estimated using IV 2SLS.
6. Clustered standard errors on firm level in parentheses: *** p<0.01, ** p<0.05, * p<0.1.

## 4.3 Heterogeneity Across Product Types

To further unpack the underlying mechanisms, we examine heterogeneous effects of diversification across product categories. Figure 4 reports the average marginal effects of modality diversification with 95% confidence intervals. While the effects are generally positive, their magnitude varies substantially across categories, with some (e.g., Automotive Parts/Accessories and Hardware/Home Improvement) exhibiting stronger effects and others showing weaker or insignificant estimates. This heterogeneity suggests that the effectiveness of modality diversification depends on product characteristics. Consistent with this interpretation, Appendix A.6 shows that the benefits of diversification are more pronounced for experiential (relative to search) and hedonic (relative to utilitarian) products, and are stronger for firms with younger and higher-income customer bases. These results are consistent with the expectation that experience, and hedonic goods rely on sensory and imagery cues that are particularly likely to benefit from cross modality exposure across multiple social media platforms. Young users and high income individuals are similarly more prone to respond to richer digital content (Bolton et al., 2013), and thus more likely to benefit from platform diversification. Please see detailed discussion in Appendix A.6.

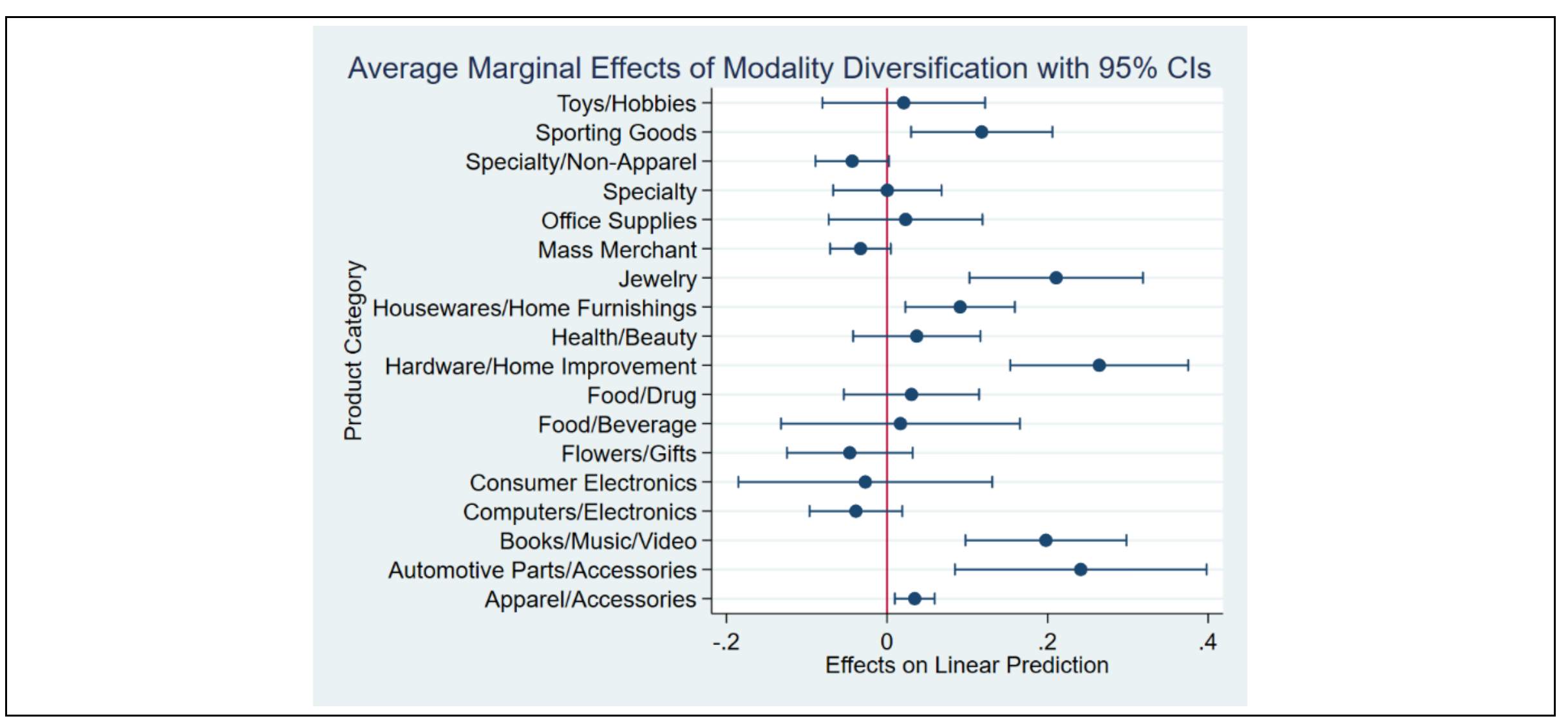

**Figure 4. Average Marginal Effects of Modality Diversification on Different Product Categories with 95% Confidence Intervals**

### 4.4 Mechanism: Diversification Improves Conversion Rather Than Traffic Volume

We next examine the mechanism through which diversified social media engagement improves firm performance. Specifically, we decompose web sales into traffic, ticket size, and conversion rate to assess whether diversification operates by attracting more visitors, increasing spending per transaction, or improving conversion. Table 5 reports the results.

Columns (2) and (3) show that engagement diversification has no significant effect on total traffic or ticket size, suggesting that diversification does not primarily expand audience size or spending per purchase. In contrast, Column (4) shows a positive and significant effect on conversion rates, indicating that diversification improves the likelihood that visitors convert into customers. These results remain robust under the 2SLS specifications in Columns (5) and (6). Columns (7)–(9) further introduce modality diversification and its interaction with engagement diversification. The interaction term is positive and significant across OLS and IV specifications, suggesting that the conversion-rate effect is stronger when engagement spans heterogeneous content modalities, consistent with Table 4.

Taken together, these findings indicate that the benefits of diversification arise primarily through improved conversion rather than increased traffic or transaction size. This mechanism is consistent with the idea that cross-platform, cross-modality engagement reinforces product exposure and facilitates purchase decisions. Additional analyses in Appendix A.7 and A.8 show that diversification reshapes traffic composition and improves traffic quality, rather than increasing overall traffic volume.

**Table 5. Mechanism Test: Diversification Increases Web Sales Through Improving Conversion Rate**

| Model | OLS | OLS | OLS | OLS | Policy IV 2SLS | Hausman IV 2SLS | OLS | Policy IV 2SLS | Hausman IV 2SLS |
|---|---|---|---|---|---|---|---|---|---|
| | (1) | (2) | (3) | (4) | (5) | (6) | (7) | (8) | (9) |
| Dependent Variable | Log web sales | Log total traffic | Log ticket size | Log conversion rate | | | | | |
| | | | | | | | | | |
| Engagement Diversification | 0.035*** | 0.028 | 0.005 | 0.0004*** | **0.0006**** | **0.0004***** | -0.0001 | **-0.0021*** | **-0.0003** |
| | (0.010) | (0.141) | (0.007) | (0.0001) | **(0.0003)** | **(0.0001)** | (0.0002) | **(0.0012)** | **(0.0002)** |
| Modality Diversification | | | | | | | 0.0006*** | **0.0024**** | **0.0008***** |
| | | | | | | | (0.0001) | **(0.0011)** | **(0.0002)** |
| Modality Diversification * Engagement Diversification | | | | | | | 0.0003*** | **0.0019**** | **0.0004***** |
| | | | | | | | (0.0001) | **(0.0009)** | **(0.0001)** |
| Other Controls | Share of Facebook/Twitter/Instagram/YouTube Engagement, Platform Number, Log Total Number of Followers, Log Total Number of Postings, Log Total Number of Engagements, Company Age, Number of SKUs, Retailer FE, Year FE, State-Year FE, Merchant Type, Product Category, Product Category-Year FE | | | | | | | | |
| Observations | 6,994 | 6,994 | 6,994 | 6,994 | 6,994 | 6,994 | 6,994 | 6,994 | 6,994 |
| R Squared | 0.971 | 0.914 | 0.959 | 0.970 | \ | \ | 0.971 | \ | \ |

Notes:

1. The dependent variable in Column (1) is the log total web sales. The dependent variable in Column (2) is the log total traffic (i.e., the number of visits to the firm website). The dependent variable in Column (3) is the log of ticket size (i.e., the average dollar amount users spend per transaction). The dependent variable in Columns (4) – (9) is the log conversion rate (i.e., the percentage of users who make a transaction out of the total number of visitors). By definition, total web sales = total traffic × ticket size × conversion rate.
2. Engagement Diversification is constructed using the same method as in Tables 3 to 5. Modality Diversification is constructed using the same method as in Tables 4 and 5. Share of Facebook/Twitter/Instagram/YouTube Engagement is constructed using the same method as in Tables 3 to 5.
3. Columns (1) – (4) and (7) report the OLS results with company and year fixed effects. Columns (5) and (8) report the 2SLS results using platform policies as instrumental variables. Columns (6) and (9) report the 2SLS results using Hausman-type instruments as instrumental variables. The first-stage F-statistic is greater than 20. Variables in bold are treated as endogenous and estimated using IV 2SLS.
4. Clustered standard errors on firm level in parentheses: *** $p<0.01$, ** $p<0.05$, * $p<0.1$.

## 4.5 Overlapping Followers

One central assumption we made on the positive impact of social media diversification is that users tend to multi-home. The main benefit comes from presenting a variety of product exposures to an overlapping set of users. To show that this is indeed the case, we attempt to measure overlapping users between platforms. To do this, we collected the full list of follower names in chronological order for each retailer on Twitter and Pinterest using public APIs in 2021. Among the five major platforms in our dataset, only Twitter and Pinterest publicly disclose follower identities, which enables cross-platform matching. For each retailer-year observation, we reconstruct the follower base corresponding to that year by trimming the historical follower list to match the reported follower counts in that year.[16] We identify two follower IDs as belonging to the same individual if the Jaro–Winkler string distance between the two IDs is below 10 percent.[17] Based on this matching procedure, we construct the **Overlapping Index** as: $OverlappingIndex_{i,t} = \frac{\#SamePerson_{i,t}}{\#TwitterFollower_{i,t} + \#PinterestFollower_{i,t}}$. Because follower identity information is available only for Twitter and Pinterest, we recompute the Engagement Diversification index using these two platforms for the overlap analysis. This measure allows us to assess whether diversification generates stronger performance effects when it is reinforced among overlapping users, consistent with a memory reinforcement mechanism. In the two-platform setting of

---

[16] For example, if retailer A had one million followers on Twitter in 2016, we approximate its 2016 follower base by selecting the earliest one million followers observed in the 2021 follower list, based on chronological ordering. This reconstruction assumes that follower accumulation is largely monotonic over time. To account for potential follower churn (e.g., unfollowing behavior), we conduct robustness checks by retaining only the earliest 90% and 80% of followers in the reconstructed set. Our main results remain qualitatively and quantitatively similar under these alternative specifications.

[17] This matching procedure assumes that the same individual is likely to use similar account names across different platforms. For instance, the Jaro–Winkler distance between "@tomcruise" and "@tom_cruise" is 6%, which falls below our baseline threshold of 10%. To ensure that our results are not sensitive to the choice of matching threshold, we conduct robustness checks using alternative cutoffs of 5% and 15%. The estimated coefficients remain qualitatively and quantitatively similar across these specifications.

Twitter and Pinterest, engagement diversification directly reflects the degree of cross-modality exposure. Therefore, if cognitive reinforcement underlies the diversification effect, we expect the effect of diversification to be more pronounced when there are more overlapping followers across the two platforms. Conversely, when audience overlap is limited, diversification may expose different users to isolated pieces of information, thereby constraining opportunities for repeated exposure and reducing its potential benefits. We further address the possibility that overlapping followers may simply proxy for loyal or highly engaged customers. To mitigate this concern, we construct the **Average Visits per User** measure, defined as total annual website visits divided by the number of unique visitors. This metric captures repeat visitation intensity and serves as a proxy for behavioral loyalty in online environments (Oliver, 1999). In digital settings, visit frequency and repeat browsing behavior are commonly interpreted as indicators of customer engagement and retention (Moe & Fader, 2004). By controlling for this measure, we distinguish the reinforcement effect of cross-platform exposure from baseline customer loyalty.

Table 6 reports the results. Columns (1)–(3) show that both engagement diversification and follower overlap are positively associated with web sales, and that their interaction is positive and significant. The main effect of diversification becomes insignificant once the interaction is included, indicating that the benefits of diversification depend critically on overlapping exposure. These results remain robust under the 2SLS specifications in Columns (4) and (5). Particularly, the 2SLS estimates in Column (4) imply that the marginal effect of diversification is given by $-0.287 + 0.495 \times$ overlap, highlighting that overlap serves as a boundary condition for the effectiveness of diversification. When overlap is low (e.g., below the threshold of 0.58 (in standardized units)), diversification tends to expose different users to isolated pieces of information, limiting opportunities for repeated exposure and memory reinforcement. In this region, the costs of coordinating content across platforms and the dilution of audience attention may outweigh the benefits, resulting in weak or even negative effects. In contrast, when overlap is sufficiently high, repeated and complementary exposures to the same users reinforce prior impressions, leading to positive and economically meaningful effects. We also show in Appendix A.9 that these results are not driven by outliers, using a winsorized sample in the regression. Columns (6)–(10) replicate the analysis using conversion rate as the dependent variable and yield similar patterns. Taken together, despite potential measurement limitations, the results are consistent with a memory-reinforcement mechanism. Diversification enhances performance primarily when it generates overlapping exposure to the same users across platforms. Given the generally high degree of user overlap across social media platforms, the benefits of diversification are likely to outweigh its coordination costs.

**Table 6. Overlapping Audience and Repeated Impression Effect**

| Model | OLS | OLS | OLS | Policy IV 2SLS | Hausman IV 2SLS | OLS | OLS | OLS | Policy IV 2SLS | Hausman IV 2SLS |
|---|---|---|---|---|---|---|---|---|---|---|
| | (1) | (2) | (3) | (4) | (5) | (6) | (7) | (8) | (9) | (10) |
| Dependent Variable | Log web sales | | | | | Log conversion rate | | | | |
| | | | | | | | | | | |
| Engagement Diversification (On Twitter and Pinterest) | 0.042*** | 0.019* | -0.005 | -0.287** | -0.063 | 0.0005*** | 0.0002* | -0.0001 | -0.0039** | -0.0007 |
| | (0.010) | (0.011) | (0.012) | (0.116) | (0.042) | (0.0001) | (0.0001) | (0.0002) | (0.0016) | (0.0005) |
| Overlapping | | 0.090*** | 0.132*** | 0.629*** | 0.235*** | | 0.0011*** | 0.0017*** | 0.0084*** | 0.0028*** |
| | | (0.015) | (0.019) | (0.204) | (0.072) | | (0.0002) | (0.0002) | (0.0027) | (0.0009) |

| Engagement Diversification * Overlapping | | | 0.038*** | **0.495*** ** | **0.133** ** | | | 0.0005*** | **0.0067*** ** | **0.0016*** |
|---|---|---|---|---|---|---|---|---|---|---|
| | | | (0.011) | **(0.187)** | **(0.065)** | | | (0.0001) | **(0.0025)** | **(0.0008)** |
| Average Visits Per User | 0.019*** | 0.018*** | 0.020*** | 0.019*** | 0.019*** | 0.0023*** | 0.0022*** | 0.0022*** | 0.0022*** | 0.0022*** |
| | (0.003) | (0.003) | (0.003) | (0.003) | (0.003) | (0.0005) | (0.0005) | (0.0006) | (0.0005) | (0.0004) |
| Other Controls | Log Number of Twitter Engagement, Log Number of Pinterest Engagement, Platform Number, Log Total Number of Followers, Log Total Number of Postings, Company Age, Number of SKUs, Retailer FE, Year FE, State-Year FE, Merchant Type, Product Category, Product Category-Year FE | | | | | | | | | |
| Observations | 6,751 | 6,751 | 6,751 | 6,751 | 6,751 | 6,751 | 6,751 | 6,751 | 6,751 | 6,751 |
| R Squared | 0.970 | 0.970 | 0.970 | \ | \ | 0.970 | 0.970 | 0.970 | \ | \ |

Notes:

1. The dependent variable in Columns (1) – (5) is the log total web sales of the focal firm in the focal year. The dependent variable in Columns (6) – (10) is the log conversion rate of the focal firm in the focal year. The sample size is slightly reduced due to the exclusion of observations without either Twitter or Pinterest.

2. Engagement Diversification is constructed as one minus the HHI of the distribution of likes received across the 2 social platforms (Twitter and Pinterest) each year. To ensure comparability across platforms, we standardize the distribution of likes on each platform by dividing it by the platform-year median before computing the HHI. This adjustment reduces distortions arising from differences in platform size and engagement intensity.

3. Overlapping is constructed as $OverlappingIndex_{i,t} = \frac{\#SamePerson_{i,t}}{\#TwitterFollower_{i,t}+\#PinterestFollower_{i,t}}$, where for each company and year, we obtain the full follower lists on Twitter and Pinterest, and treat 2 follower IDs as the same person if their Jaro-Winkler distance is below 10%. Then we calculate the proportion of the same person over the total number of followers on both platforms as the overlapping index.

4. Average visits per user is constructed as the ratio of the total number of web visits to the number of unique visitors.

5. Columns (1) – (3) and (6) – (8) report the OLS results with company and year fixed effects. Columns (4) and (9) report the 2SLS results using platform policies as instrumental variables. Columns (5) and (10) report the 2SLS results using Hausman-type instruments as instrumental variables. The first-stage F-statistic is greater than 20. Variables in bold are treated as endogenous and estimated using IV 2SLS.

6. Clustered standard errors on firm level in parentheses: *** p<0.01, ** p<0.05, * p<0.1.

# 5. Discussion

Our empirical results demonstrate that a diversified social media engagement strategy enhances e-commerce performance primarily through cross-modality complementarity. Firms that distribute engagement across mixed, image, and video modalities experience higher conversion efficiency and improved traffic quality, which ultimately translates into higher web sales. The benefits of diversification are particularly pronounced for experience goods, hedonic products, younger customer segments, and higher-income customers—contexts in which rich sensory cues and affective processing are more influential.

To mitigate potential endogeneity concerns arising from time-varying firm characteristics and category-level shocks, we employ multiple identification strategies, including panel data methods, policy changes on social platforms, Hausman-type instrumental variables, and additional robustness analyses. The first-stage statistics indicate strong instrument relevance, and results remain consistent across alternative diversification measures and subsamples.

This study has several limitations. First, our sample focuses on leading U.S. e-commerce retailers and major social media platforms. Smaller firms or emerging platforms may exhibit different dynamics. Second, although we provide evidence consistent with cognitive reinforcement and memory mechanisms, future research could directly examine consumer-level behavioral data to further validate these micro-foundations. Third, platform algorithms and content formats evolve over time, and future studies may explore how algorithmic changes interact with diversification strategies. Finally, future research could benefit from access to firm-level data on platform-specific marketing budgets or expenditures, which would allow for a more precise assessment of how resource allocation shapes diversification strategies and their performance implications.

Despite these limitations, our study offers important contributions to both academic research and managerial practice. First, while prior research has largely examined single-platform strategies or platform-specific effects, our study shifts attention to cross-platform engagement structure and demonstrates that allocation patterns across platforms matter independently of individual platform performance. Second, we provide large-scale empirical evidence linking multi-platform engagement diversification to firm-level sales performance through conversion and traffic-quality mechanisms. Finally, our findings suggest that firms should move beyond platform adoption breadth and instead strategically design engagement portfolios that leverage complementarities across content modalities and audience overlap. In increasingly fragmented digital environments, performance gains stem not from platform expansion alone, but from coordinated and cognitively reinforcing engagement structures.

# Appendix

### A.1 Social Media Data Extraction through Wayback Machine

In this section, we demonstrate the detailed process of social media data extraction through the Wayback Machine. From the Internet Retailer Guide, we obtain each company's official web domain (URLs). We then obtain each company's official social media handle through its official web domain and use the Wayback Machine to download the web snapshots of these social media handles from 2012 to 2019. In each year we download the earliest and latest 5 snapshots. An example of snapshots is shown below:

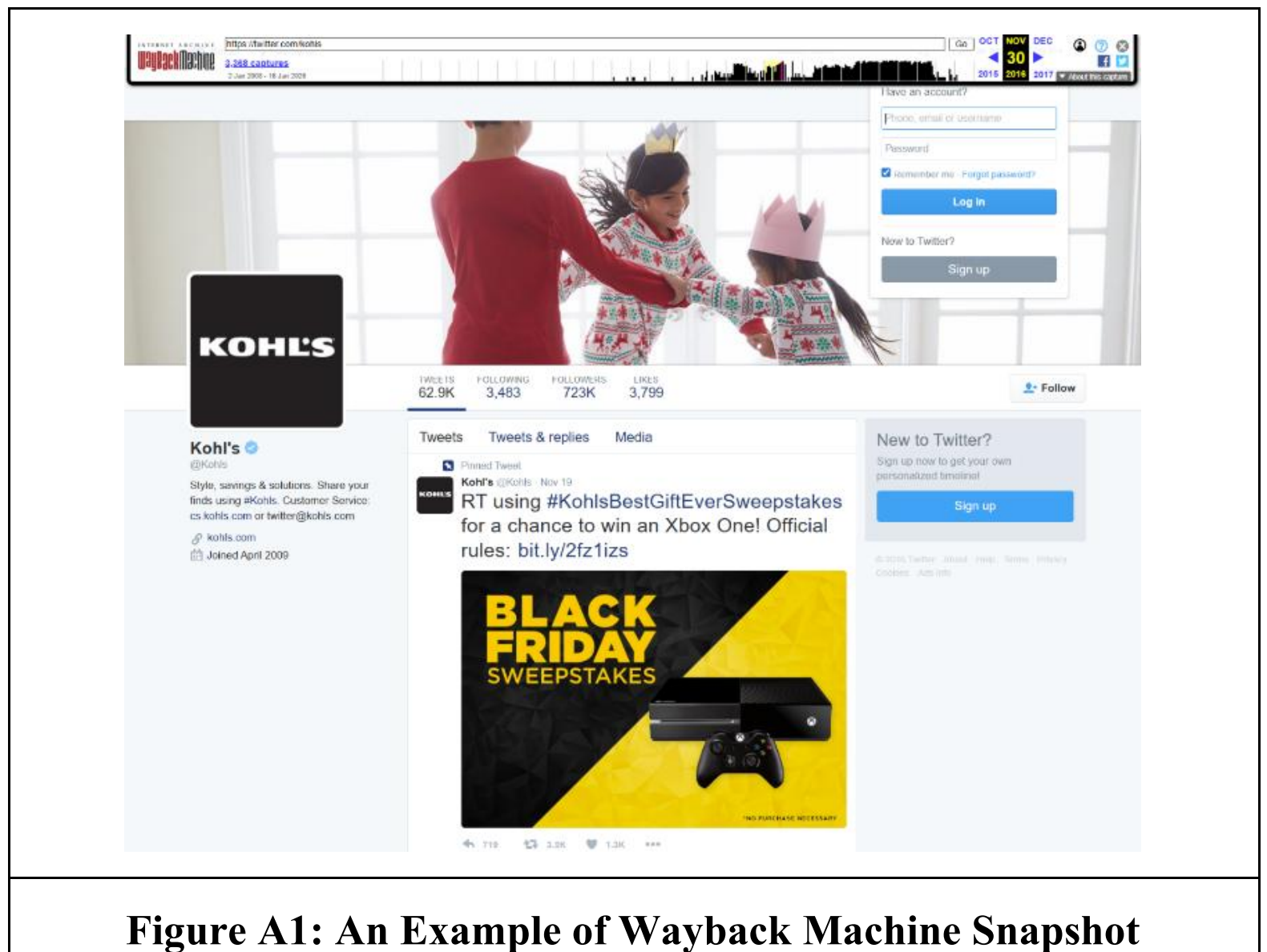

**Figure A1: An Example of Wayback Machine Snapshot**

Because our sample consists of the top 1,000 U.S. e-commerce retailers, all firms maintain an observable digital footprint, and their official social media pages are consistently archived in the Wayback Machine. This comprehensive archival coverage enables us to systematically reconstruct engagement activity across firms and years. From these snapshots, we extract the total numbers of followers, posts, and likes. We compute annual posting and like counts by taking the difference between the cumulative counts at the end of year t and the end of year t–1. Because the metrics are cumulative and mechanically increasing over time, measurement error due to minor timing mismatches in archived snapshots is unlikely to systematically bias our annual differences. When a year-end snapshot was unavailable, we identified the closest archived snapshot within a short time window and manually updated the cumulative metrics. In a small number of cases, archived snapshots did not display cumulative statistics (e.g., total posts or total likes/pins) due to platform interface changes or incomplete captures. For these retailer–platform–year observations, we implemented a predefined fallback procedure: we manually collected all posts published by the retailer on the focal platform within the calendar year and aggregated the corresponding engagement counts (likes/saves, depending on platform conventions). This post-level reconstruction was used only when the cumulative year-end metrics were unavailable, and we applied the same coding rules across firms and years to ensure consistency.

## A.2 Robustness Tests for the Measurement of Diversification

### A.2.1 Alternative Measurement of Diversification

In this section, we show that the main results of this paper are robust across different specifications of the engagement diversification measure. In the following table, we show that the regression results are consistent using the Engagement Diversification, Audience Diversification, or Content Diversification as the independent variable. The **Audience Diversification** is calculated as one minus the HHI of the distribution of followers across social media platforms: $Audience_Diversification_{i,t} = 1 - FollowerHHI_{i,t} = 1 - \sum_{p=1}^{5} follower_share_{i,t,p}^{2}$ where $follower_share_{i,t,p}$ stands for the share of the number of followers on the social platform $p$ for company $i$ in year $t$. For example, Facebook's share for the company $i$ in year $t$ is: $follower_share_{i,t,p=Facebook} = \frac{\#\ Facebook\ Followers_{i,t}}{\sum_{All\ platforms} \#\ Followers_{i,t}}$. The **Content Diversification** is computed as one minus the HHI of the annual number of posts across platforms: $Content_Diversification_{i,t} = 1 - PostingHHI_{i,t} = 1 - \sum_{p=1}^{5} posting_share_{i,t,p}^{2}$ where $posting_share_{i,t,p}$ stands for the share of the number of posts on the social platform $p$ published by company $i$ in year $t$. For example, Facebook's share for the company $i$ in year $t$ is: $posting_share_{i,t,p=Facebook} = \frac{\#\ Posts\ on\ Facebook_{i,t}}{\sum_{All\ platforms} \#\ Posts_{i,t}}$. Columns (1), (4), and (7) report the OLS results with company and year fixed effects. Columns (2), (5), and (8) report the 2SLS results using the platform policy shocks as instrumental variables. Columns (3), (6), and (9) report the 2SLS results using the Hausman-type instrument as instrumental variables. The first-stage F-statistic is greater than 20 in each 2SLS result. Variables in bold are treated as endogenous and estimated using IV 2SLS. All columns show positive, statistically significant coefficients. The consistency of the sign and statistical significance across these distinct identification strategies provides reassurance that the estimated relationship is not driven by a particular source of variation. Note the difference in coefficient magnitude across the two instrumental variable strategies is expected. The platform policy instruments capture exogenous shocks to platform algorithms that primarily affect firms highly engaged with social media platforms, while the Hausman-type instruments capture peer-driven diffusion of social media strategies within geographic regions. As a result, the two strategies identify local average treatment effects for different subsets of firms. The policy shocks likely affect firms with stronger social media capabilities and higher platform dependence, leading to larger estimated effects, whereas the Hausman instruments capture more gradual imitation among firms with weaker engagement strategies, resulting in smaller effect sizes.

**Table A.2.1 Robust Results using Different Diversification Metrics**

| Model | OLS | Policy IV 2SLS | Hausman IV 2SLS | OLS | Policy IV 2SLS | Hausman IV 2SLS | OLS | Policy IV 2SLS | Hausman IV 2SLS |
|---|---|---|---|---|---|---|---|---|---|
| DV: Log web sales | (1) | (2) | (3) | (4) | (5) | (6) | (7) | (8) | (9) |
| | | | | | | | | | |
| Engagement Diversification | 0.043*** | **0.074*** ** | **0.033** ** | | | | | | |
| | (0.010) | **(0.020)** | **(0.013)** | | | | | | |
| Audience Diversification | | | | 0.035** | **0.083*** ** | **0.026** ** | | | |

| | | | | (0.016) | **(0.023)** | **(0.012)** | | | |
|---|---|---|---|---|---|---|---|---|---|
| Content Diversification | | | | | | | 0.028*** | **0.084***** | **0.059***** |
| | | | | | | | (0.009) | **(0.024)** | **(0.013)** |
| Other Controls | Platform Number, Log Total Number of Followers, Log Total Number of Postings, Log Total Number of Engagements, Company Age, Number of SKUs, Retailer FE, Year FE, State-Year FE, Merchant Type, Product Category | | | | | | | | |
| Observations | 6,994 | 6,994 | 6,994 | 6,994 | 6,994 | 6,994 | 6,994 | 6,994 | 6,994 |
| R Squared | 0.970 | \ | \ | 0.970 | \ | \ | 0.970 | \ | \ |

Notes:
1. The dependent variable is the log total web sales of the focal firm in the focal year.
2. Three diversification indexes are used. The Engagement Diversification is constructed as one minus the HHI of the distribution of likes received across platforms in a given year. Higher values indicate that user engagement is broadly distributed rather than concentrated on a single platform. The Audience Diversification is calculated as one minus the Herfindahl–Hirschman Index (HHI) of the distribution of followers across social media platforms. Higher values indicate that a firm's audience base is more evenly distributed across platforms rather than concentrated on a single platform. The Content Diversification is computed as one minus the HHI of the annual number of posts across platforms. This measure captures the extent to which a firm spreads its content production efforts across multiple platforms. To ensure comparability across platforms, we standardize each metric by dividing it by the platform-year median before computing the HHI. This adjustment reduces distortions arising from differences in platform size and engagement intensity. All three metrics are standardized for the interpretability of the results.
3. Platform Number measures how many social platforms (among Facebook, Twitter, Instagram, YouTube, Pinterest, value range 0~5) the focal retailer has adopted in the given year.
4. Columns (1), (4), and (7) report the OLS results with company and year fixed effects. Columns (2), (5), and (8) report the 2SLS results using the platform policy shocks as instrumental variables. Columns (3), (6), and (9) report the 2SLS results using the Hausman-type instrument as instrumental variables. The first-stage F-statistic is greater than 20 in each 2SLS result. Variables in bold are treated as endogenous and estimated using IV 2SLS.
5. Clustered standard errors on firm level in parentheses: *** p<0.01, ** p<0.05, * p<0.1.

### A.2.2 Substitute HHI using Gini Coefficient

We further replicate our baseline results using the Gini Coefficient rather than the HHI to calculate diversification levels. The Gini Coefficient is also widely used to measure the level of inequality in the allocation across different subjects. Gini Coefficient is calculated as: $Gini_Coefficient_{i,t} = \frac{\sum_{j=1}^{n}\sum_{k=1}^{n}|x_j - x_k|}{2n^2\bar{x}}$, where $Gini_Coefficient_{i,t}$ stands for the Gini Coefficient of the company $i$ at time $t$, $x_j$ and $x_k$ represent the number of likes on the platform $j$ and platform $k$ after standardization using the platform-year median, $n$ stands for the platform number, which is 5 in our data, and $\bar{x}$ stands for the average number of likes across all platforms. Similar to HHI, the Gini Coefficient also ranges from 0 to 1. A higher Gini Coefficient suggests that the company is more concentrated on some social platforms than others. Next, we calculate the engagement diversification metric using 1 minus the Gini Coefficient. For interpretability purposes, we also standardize this diversification index by subtracting its sample mean and dividing by its sample standard deviation. The following table shows the summary statistics of Gini Coefficient:

**Table A.2.2 Summary Statistics for the Gini Coefficient (n=6,994)**

| Statistic | Mean | St. Dev. | Min | Max |
|---|---|---|---|---|
| Engagement Diversification (based on Gini Coefficient) | 0 | 1.00 | -2.59 | 1.31 |

The following table replicates the baseline results in Table 3 using the same empirical models while substituting the engagement diversification measure using the one calculated based on the Gini Coefficient. The results are consistent.

**Table A.2.3 Robust Results using Gini Coefficient**

| Model | OLS | OLS | Policy IV 2SLS | Hausman IV 2SLS |
|---|---|---|---|---|
| DV: Log web sales | (1) | (2) | (3) | (4) |
| | | | | |
| Engagement Diversification (based on Gini Coefficient) | 0.023*** | 0.019** | **0.024**** | **0.018**** |
| | (0.008) | (0.010) | **(0.011)** | **(0.010)** |
| Other Controls | Share of Engagement on Facebook, Twitter, Instagram, and YouTube, Platform Number, Log Total Number of Followers, Log Total Number of Postings, Log Total Number of Engagements, Company Age, Number of SKUs, Retailer FE, Year FE, State-Year FE, Merchant Type, Product Category | | | |
| Observations | 6,994 | 6,994 | 6,994 | 6,994 |

| R Squared | 0.970 | 0.971 | \ | \ |
|---|---|---|---|---|

Notes:
1. The dependent variable is the log total web sales of the focal firm in the focal year.
2. The Engagement Diversification is constructed as one minus the Gini Coefficient of the distribution of likes received across platforms in a given year. Higher values indicate that user engagement is broadly distributed rather than concentrated on a single platform.
3. Share of Facebook/Twitter/Instagram/YouTube Engagement is calculated as the ratio of the number of likes received on a single platform over the total number of likes received from all platforms in a given year. To ensure comparability across platforms, we standardize each metric by dividing it by the platform-year median before computing the share ratio. This adjustment reduces distortions arising from differences in platform size and engagement intensity. The Share of Pinterest Engagement has dropped because of multicollinearity (all shares add up to 1).
4. Platform Number measures how many social platforms (among Facebook, Twitter, Instagram, YouTube, Pinterest, value range 0~5) that the focal retailer has adopted in the given year.
5. Columns (1) and (2) report the OLS results with company and year fixed effects. Column (3) reports the 2SLS results using platform policies as instrumental variables. Column (4) reports the 2SLS results using Hausman-type instruments as instrumental variables. The first-stage F-statistic is greater than 20. Variables in bold are treated as endogenous and estimated using IV 2SLS.
6. Clustered standard errors on firm level in parentheses: *** p<0.01, ** p<0.05, * p<0.1.

### A2.3 Matching for Robustness

We further show that the results are robust using matched samples to run regressions. We follow Blackwell et al. (2009) to conduct coarsened exact matching on the sample data, and then use the matched data to run regression models. The results are shown below. The estimated coefficients are qualitatively unchanged.

**Table A.2.4 Robust Results using Coarsened Exact Matching**

| Model | OLS | CEM Matching + OLS | CEM Matching + Policy IV 2SLS | CEM Matching + Hausman IV 2SLS |
|---|---|---|---|---|
| DV: Log web sales | (1) | (2) | (3) | (4) |
| | | | | |
| Engagement Diversification | 0.043*** | 0.049*** | **0.035***** | **0.044***** |
| | (0.010) | (0.009) | **(0.010)** | **(0.011)** |
| Other Controls | Share of Engagement on Facebook, Twitter, Instagram, and YouTube, Platform Number, Log Total Number of Followers, Log Total Number of Postings, Log Total Number of Engagements, Company Age, Number of SKUs, Retailer FE, Year FE, State-Year FE, Merchant Type, Product Category | | | |
| Observations | 6,994 | 5,126 | 5,126 | 5,126 |
| R Squared | 0.970 | 0.972 | | |

Notes:
1. The dependent variable is the log total web sales of the focal firm in the focal year.
2. The Engagement Diversification is constructed as one minus the HHI of the distribution of likes received across platforms in a given year. Higher values indicate that user engagement is broadly distributed rather than concentrated on a single platform.
3. Share of Facebook/Twitter/Instagram/YouTube Engagement is calculated as the ratio of the number of likes received on a single platform over the total number of likes received from all platforms in a given year. To ensure comparability across platforms, we standardize each metric by dividing it by the platform-year median before computing the share ratio. This adjustment reduces distortions arising from differences in platform size and engagement intensity. The Share of Pinterest Engagement has dropped because of multicollinearity (all shares add up to 1).
4. Platform Number measures how many social platforms (among Facebook, Twitter, Instagram, YouTube, Pinterest, value range 0~5) that the focal retailer has adopted in the given year.
5. Columns (1) reports the OLS results with company and year fixed effects. Column (2)-(4) use coarsened exact matching to construct a matched sample and then use the matched sample to run the regression. Column (2) reports the OLS results with fixed effects. Column (3) reports the 2SLS results using platform policies as instrumental variables. Column (4) reports the 2SLS results using Hausman-type instruments as instrumental variables. The first-stage F-statistic is greater than 20. Variables in bold are treated as endogenous and estimated using IV 2SLS.
6. Clustered standard errors on firm level in parentheses: *** p<0.01, ** p<0.05, * p<0.1.

### A.3 Robustness Tests for the Platform Policy Change Instruments

In this section, we conduct robustness tests regarding the policy changes instruments. In the following table, we show the first stage regression results of the platform policy changes instruments. The dependent variable is the focal firm's Engagement Diversification in the following year. The independent variables are the policy change instrumental variables used in the main text: The YouTube Red policy change in 2015, Facebook News Feed Policy Change in 2018, and Pinterest Buyable Pins Policy Change in 2015.

Column (1) shows the result of YouTube Red Policy Change using the full sample. The estimated coefficient is not statistically significant. Such null results provide useful placebo tests and further support the validity of our identification strategy. In particular, they help rule out the possibility that our findings are driven by general time trends in social media marketing or by coincidental platform policy changes during the sample period. Instead, the results are consistent with our mechanism that only platform policies that meaningfully alter the relative attractiveness of platforms generate shifts in firms' engagement allocation. Columns (2) and (3) show the results of Facebook News Feed Policy Change using the sample of companies with high versus low Facebook exposure in 2018 (separated at the median), respectively. The results show that this policy significantly increased engagement diversification only among companies with high Facebook exposure, whereas it had no significant effect among companies with low Facebook exposure. Columns (4) and (5) show the results of Pinterest Buyable Pins Policy Change using the sample of companies with high/low Pinterest exposure in 2015 (separated at the median), respectively. Similarly, the results show that this policy significantly decreased engagement diversification only among companies with high Pinterest exposure, whereas it had no significant effect among companies with low Pinterest exposure.

Taken together, these results provide additional support for the relevance of the policy-based instruments. The policy shocks significantly affect engagement diversification only among firms with greater prior exposure to the affected platform, where the economic impact of the policy change is expected to be strongest. At the same time, the absence of significant effects among low-exposure firms, together with the null result for the YouTube Red policy, suggests that our findings are unlikely to be driven by broad time trends in social media marketing or by general platform dynamics during the sample period. Instead, the results are consistent with our identification strategy that relies on platform-specific shocks that differentially affect firms depending on their prior platform exposure.

**Table A.3.1 Diversification Effects Beyond Individual Platform Effects**

| Model | OLS | OLS | OLS | OLS | OLS |
|---|---|---|---|---|---|
| Sample | Full Sample | High Facebook Exposure | Low Facebook Exposure | High Pinterest Exposure | Low Pinterest Exposure |
| DV: Lead Engagement Diversification | (1) | (2) | (3) | (4) | (5) |
| | | | | | |
| YouTube Red Policy Change | 0.024 | | | | |

| | (0.162) | | | | |
|---|---|---|---|---|---|
| Facebook News Feed Policy Change | | 0.233*** | -0.141 | | |
| | | (0.062) | (0.253) | | |
| Pinterest Buyable Pins Policy Change | | | | -0.175*** | 0.050 |
| | | | | (0.049) | (0.134) |
| Controls | Share of Engagement on Facebook, Twitter, Instagram, and YouTube, Platform Number, Total Number of Followers, Total Number of Posts, Total Number of Engagements, Company Age, Number of SKUs, Retailer FE, Year FE, State-Year FE, Merchant Type, Product Category | | | | |
| Observations | 6,994 | 3,020 | 3,974 | 3,124 | 3,870 |
| R Squared | 0.393 | 0.621 | 0.414 | 0.614 | 0.518 |

Notes:
1. The dependent variable is the Engagement Diversification of the focal firm in the following year.
2. The independent variables are the policy change instrumental variables used in the main text: The YouTube Red policy change in 2015, Facebook News Feed Policy Change in 2018, and Pinterest Buyable Pins Policy Change in 2015.
3. Column (1) shows the result of YouTube Red Policy Change using the full sample. Columns (2) and (3) show the results of Facebook News Feed Policy Change using the sample of companies with high/low Facebook exposure in 2018 (separated at the median), respectively. Columns (4) and (5) show the results of Pinterest Buyable Pins Policy Change using the sample of companies with high/low Pinterest exposure in 2015 (separated at the median), respectively.
4. Clustered standard errors on firm level in parentheses: *** p<0.01, ** p<0.05, * p<0.1

### A.4 Robustness Tests after Excluding Companies in New York and California

In this section, we show that the results in the main text Table 3 remain substantially the same after excluding the sample of companies located in the states of New York and California. Such results confirm that the main results in this paper are not driven by a few high-performing retailers that are geographically clustered.

**Table A.4.1. Social Media Engagement Diversification Effect on Web Sales**

| Model | OLS | OLS | Policy IV 2SLS | Hausman IV 2SLS |
|---|---|---|---|---|
| DV: Log web sales | (1) | (2) | (3) | (4) |
| | | | | |
| Engagement Diversification | 0.040*** | 0.036*** | **0.056**** | **0.035***** |
| | (0.010) | (0.012) | **(0.026)** | **(0.013)** |
| Other Controls | Share of Engagement on Facebook, Twitter, Instagram, and YouTube, Platform Number, Log Total Number of Followers, Log Total Number of Postings, Log Total Number of Engagements, Company Age, Number of SKUs, Retailer FE, Year FE, State-Year FE, Merchant Type, Product Category | | | |
| Observations | 4,724 | 4,724 | 4,724 | 4,724 |
| R Squared | 0.972 | 0.972 | \ | \ |

Notes:
1. The dependent variable is the log total web sales of the focal firm in the focal year.
2. The Engagement Diversification is constructed as one minus the HHI of the distribution of likes received across platforms in a given year. Higher values indicate that user engagement is broadly distributed rather than concentrated on a single platform.
3. Share of Facebook/Twitter/Instagram/YouTube Engagement is calculated as the ratio of the number of likes received on a single platform over the total number of likes received from all platforms in a given year. To ensure comparability across platforms, we standardize each metric by dividing it by the platform-year median before computing the share ratio. This adjustment reduces distortions arising from differences in platform size and engagement intensity. The Share of Pinterest Engagement has dropped because of multicollinearity (all shares add up to 1).
4. Platform Number measures how many social platforms (among Facebook, Twitter, Instagram, YouTube, Pinterest, value range 0~5) that the focal retailer has adopted in the given year.
5. Columns (1) and (2) report the OLS results with company and year fixed effects. Column (3) reports the 2SLS results using platform policies as instrumental variables. Column (4) reports the 2SLS results using Hausman-type instruments as instrumental variables. The first-stage F-statistic is greater than 20. Variables in bold are treated as endogenous and estimated using IV 2SLS.
6. Clustered standard errors on firm level in parentheses: *** p<0.01, ** p<0.05, * p<0.1.

## A.5 Correlation Matrix for Main Variables

### Table A.5 Correlation Matrix for Main Variables

| Variables | (1) | (2) | (3) | (4) | (5) | (6) | (7) | (8) | (9) | (10) | (11) | (12) | (13) | (14) |
|---|---|---|---|---|---|---|---|---|---|---|---|---|---|---|
| (1) Log Web Sales (dollars) | 1.000 | | | | | | | | | | | | | |
| (2) Log Conversion Rate | 0.210 | 1.000 | | | | | | | | | | | | |
| (3) Log Total Traffic | 0.614 | -0.011 | 1.000 | | | | | | | | | | | |
| (4) Log Ticket Size (dollars) | 0.084 | -0.398 | -0.183 | 1.000 | | | | | | | | | | |
| (5) Engagement Diversification | -0.092 | -0.088 | -0.113 | -0.037 | 1.000 | | | | | | | | | |
| (6) Audience Diversification | 0.066 | 0.069 | 0.043 | -0.088 | 0.921 | 1.000 | | | | | | | | |
| (7) Content Diversification | -0.020 | -0.016 | -0.050 | -0.068 | 0.911 | 0.956 | 1.000 | | | | | | | |
| (8) Modality Diversification | 0.390 | 0.388 | 0.296 | -0.050 | 0.362 | 0.544 | 0.472 | 1.000 | | | | | | |
| (9) Cross-Mode Complementarity | 0.315 | 0.314 | 0.211 | -0.033 | 0.375 | 0.529 | 0.540 | 0.959 | 1.000 | | | | | |
| (10) Same-Mode Complementarity | -0.289 | -0.281 | -0.316 | 0.117 | 0.337 | -0.009 | 0.055 | -0.278 | -0.222 | 1.000 | | | | |
| (11) has Facebook | 0.004 | 0.003 | -0.006 | -0.035 | 0.026 | 0.054 | 0.055 | 0.035 | 0.038 | -0.072 | 1.000 | | | |
| (12) has Twitter | 0.052 | 0.053 | 0.029 | -0.030 | 0.094 | 0.120 | 0.143 | 0.071 | 0.098 | -0.070 | 0.506 | 1.000 | | |
| (13) has YouTube | 0.173 | 0.172 | 0.112 | 0.024 | -0.034 | 0.027 | 0.017 | 0.087 | 0.079 | -0.122 | 0.268 | 0.300 | 1.000 | |
| (14) has Pinterest | 0.091 | 0.093 | 0.037 | -0.039 | 0.098 | 0.167 | 0.172 | 0.103 | 0.113 | -0.084 | 0.217 | 0.254 | 0.256 | 1.000 |
| (15) has Instagram | 0.179 | 0.183 | -0.073 | 0.009 | 0.078 | 0.134 | 0.164 | 0.148 | 0.183 | 0.006 | 0.137 | 0.155 | 0.216 | 0.395 |
| (16) Facebook Engagement Share | 0.421 | 0.419 | 0.417 | -0.131 | -0.144 | 0.176 | 0.064 | 0.441 | 0.343 | -0.638 | 0.078 | 0.070 | 0.145 | 0.139 |
| (17) Twitter Engagement Share | -0.093 | -0.097 | -0.049 | 0.013 | 0.169 | 0.080 | 0.072 | -0.129 | -0.140 | -0.033 | -0.031 | -0.009 | -0.056 | -0.083 |
| (18) Instagram Engagement Share | -0.026 | -0.016 | -0.087 | 0.046 | 0.337 | 0.211 | 0.218 | -0.009 | 0.001 | 0.561 | 0.000 | 0.005 | -0.009 | 0.067 |
| (19) YouTube Engagement Share | 0.116 | 0.108 | 0.092 | -0.013 | -0.303 | -0.312 | -0.333 | 0.026 | 0.001 | 0.042 | 0.008 | 0.011 | 0.066 | 0.040 |
| (20) Pinterest Engagement Share | -0.284 | -0.279 | -0.255 | 0.062 | 0.002 | -0.093 | -0.007 | -0.239 | -0.161 | 0.110 | -0.041 | -0.051 | -0.101 | -0.110 |
| (21) Platform Number | 0.186 | 0.189 | 0.021 | -0.014 | 0.091 | 0.177 | 0.196 | 0.167 | 0.192 | -0.092 | 0.458 | 0.520 | 0.614 | 0.745 |
| (22) Log Total Number of Followers | 0.401 | 0.394 | 0.319 | -0.060 | -0.339 | -0.182 | -0.257 | 0.187 | 0.115 | -0.250 | 0.416 | 0.377 | 0.526 | 0.414 |
| (23) Log Total Number of Postings | 0.422 | 0.422 | 0.335 | -0.118 | 0.134 | 0.271 | 0.156 | 0.333 | 0.231 | -0.165 | 0.186 | 0.293 | 0.285 | 0.343 |
| (24) Log Total Number of Engagement | 0.524 | 0.517 | 0.473 | -0.134 | -0.072 | 0.198 | 0.062 | 0.460 | 0.339 | -0.452 | 0.102 | 0.101 | 0.202 | 0.211 |
| (25) Average Visits per User | 0.023 | 0.021 | 0.164 | -0.021 | -0.009 | -0.007 | -0.015 | 0.003 | -0.005 | -0.021 | -0.011 | -0.016 | -0.046 | -0.073 |
| (26) Age | 0.281 | 0.284 | 0.094 | 0.040 | -0.087 | -0.077 | -0.080 | 0.003 | -0.001 | -0.027 | 0.024 | 0.045 | 0.078 | 0.048 |
| (27) SKUs on Web | 0.001 | 0.001 | -0.034 | 0.016 | 0.008 | 0.005 | 0.010 | -0.007 | -0.002 | 0.005 | 0.003 | 0.004 | 0.008 | 0.011 |

| | (15) | (16) | (17) | (18) | (19) | (20) | (21) | (22) | (23) | (24) | (25) | (26) | (27) |
|---|---|---|---|---|---|---|---|---|---|---|---|---|---|
| (15) has Instagram | 1.000 | | | | | | | | | | | | |
| (16) Facebook Engagement Share | 0.123 | 1.000 | | | | | | | | | | | |
| (17) Twitter Engagement Share | -0.146 | -0.073 | 1.000 | | | | | | | | | | |
| (18) Instagram Engagement Share | 0.210 | -0.081 | 0.207 | 1.000 | | | | | | | | | |
| (19) YouTube Engagement Share | 0.104 | -0.065 | -0.224 | -0.190 | 1.000 | | | | | | | | |
| (20) Pinterest Engagement Share | -0.183 | -0.544 | -0.423 | -0.481 | -0.310 | 1.000 | | | | | | | |
| (21) Platform Number | 0.744 | 0.183 | -0.126 | 0.124 | 0.089 | -0.177 | 1.000 | | | | | | |
| (22) Log Total Number of Followers | 0.438 | 0.404 | -0.202 | 0.002 | 0.342 | -0.362 | 0.663 | 1.000 | | | | | |
| (23) Log Total Number of Postings | 0.431 | 0.403 | -0.059 | 0.195 | 0.118 | -0.409 | 0.510 | 0.626 | 1.000 | | | | |
| (24) Log Total Number of Engagement | 0.272 | 0.808 | -0.246 | 0.018 | 0.150 | -0.508 | 0.309 | 0.604 | 0.624 | 1.000 | | | |
| (25) Average Visits per User | -0.069 | 0.012 | 0.018 | -0.036 | -0.006 | 0.006 | -0.081 | -0.042 | -0.033 | -0.003 | 1.000 | | |
| (26) Age | 0.094 | 0.015 | -0.019 | -0.016 | 0.082 | -0.038 | 0.099 | 0.117 | 0.067 | 0.040 | -0.002 | 1.000 | |
| (27) SKUs on Web | 0.018 | -0.013 | -0.006 | -0.009 | -0.006 | 0.020 | 0.017 | -0.015 | -0.007 | -0.012 | -0.005 | 0.011 | 1.000 |

### A.6 Heterogeneous Impact of Diversification on Product Types and Customer Segments

In this section, we further examine heterogeneous effects across product types and customer segments that differ in their reliance on cognitive processing and responsiveness to marketing cues. Specifically, we focus on experience goods versus search goods, hedonic versus utilitarian goods, as well as customer segments characterized by younger and higher-income consumers. These dimensions are theoretically relevant because the effectiveness of diversified social media engagement depends on the extent to which repeated and multimodal information can shape consumer perception, memory, and preference formation (Gu & Li, 2023). Experience goods are those whose quality is difficult to evaluate prior to consumption and therefore benefit from richer informational cues and experiential content (Nelson, 1970, 1974, Hong & Pavlou, 2014). In contrast, search goods can be evaluated based on objective attributes before purchase. Following the literature, products in the following categories are defined as experience (as opposed to search) goods: Apparel/Accessories, Books/Music/Video, Flowers/Gifts, Food/Beverage, Health/Beauty, Jewelry, Housewares/Home Furnishings, Sporting Goods, Toys/Hobbies, and Specialty/Specialty Non-Apparel. Similarly, hedonic goods are associated with affective and experiential consumption motives, whereas utilitarian goods primarily satisfy functional or instrumental needs (Batra & Ahtola, 1991; Dhar & Wertenbroch, 2000; Okada, 2005). Because experience and hedonic goods rely more heavily on imagery, affect, and sensory cues, they are particularly likely to benefit from cross-modality exposure across social media platforms. Accordingly, we define the products in the following categories as hedonic (as opposed to utilitarian) goods: Apparel/Accessories, Books/Music/Video, Flowers/Gifts, Health/Beauty, Jewelry, Housewares/Home Furnishings, and Toys/Hobbies. We also examine heterogeneity across customer demographics. Prior research suggests that younger consumers are more digitally engaged and more active across multiple platforms (Bolton et al., 2013), while higher-income consumers may exhibit distinct online consumption patterns and greater responsiveness to rich digital content (Donthu & Garcia, 1999). We therefore construct measures capturing customer composition. Specifically, we define Young Customers Ratio as the proportion of customers aged below 35, and High-Income Customers Ratio as the proportion of customers with annual income above $100,000. These data are also provided by the survey conducted by Internet Retailer Guide.

For experience goods, where product quality is difficult to evaluate prior to purchase, consumers are more likely to rely on indirect signals and repeated exposure, making them more susceptible to reinforcement effects across platforms. Similarly, hedonic goods are often associated with affective and sensory appeal, which can be more effectively conveyed through diverse content formats (e.g., images and videos), thereby enhancing the value of modality diversification. In addition, younger consumers tend to be more active on social media and more receptive to multimodal and repetitive content, increasing the likelihood that cross-platform exposure reinforces memory and influences behavior. High-income consumers, on the other hand, may exhibit lower price sensitivity and greater responsiveness to branding and experiential cues, further amplifying the impact of diversified engagement strategies.

Table A.6.1 reports the results. All interaction terms are mean-centered. Column (1) first confirms the baseline pattern: modality diversification is positively associated with web sales, and its interaction with engagement diversification is significantly positive, indicating that diversified engagement generates greater value when it spans multiple content modalities. Columns (2) and (3) examine heterogeneity across product types. The triple interaction between engagement diversification, modality diversification, and experience products is positive and significant, suggesting that cross-modality engagement is particularly beneficial for experience goods. A similar pattern emerges for hedonic products: the triple interaction with the hedonic product indicator is also positive and statistically significant. These results indicate that modality heterogeneity is more valuable when product evaluation relies more heavily on rich or affective information. Columns (4) and (5) explore customer heterogeneity. The interaction effects show that cross-modality engagement is more valuable for retailers serving younger and higher-income customers. These customer groups are more likely to process and respond to diverse multimedia content formats, which amplifies the benefits of modality diversification.[18]

---

[18] Because Table A.6.1 reports multiple heterogeneity tests around the same mechanism, we adjust the p-values of the four triple-interaction terms

Taken together, these findings suggest that the value of diversified engagement arises not merely from spreading activity across platforms, but from leveraging complementary content modalities that better match product characteristics and audience information-processing preferences.

**Table A.6.1. Cross-Modality Complementarities on Product and Customer Types**

| Model | OLS | OLS | OLS | OLS | OLS |
|---|---|---|---|---|---|
| DV: Log web sales | (1) | (2) | (3) | (4) | (5) |
| | | | | | |
| Engagement Diversification | -0.006 | -0.006 | -0.008 | -0.001 | 0.001 |
| | (0.012) | (0.013) | (0.013) | (0.013) | (0.012) |
| Modality Diversification | 0.051*** | 0.054*** | 0.056*** | 0.047*** | 0.051*** |
| | (0.010) | (0.010) | (0.010) | (0.010) | (0.010) |
| Modality Diversification * Engagement Diversification | 0.022*** | 0.018*** | -0.034* | 0.006 | 0.009 |
| | (0.005) | (0.006) | (0.018) | (0.009) | (0.007) |
| Experience Product | | 0.038 | | | |
| | | (0.083) | | | |
| Experience Product * Modality Diversification * Engagement Diversification | | 0.033*** | | | |
| | | (0.011) | | | |
| Hedonic Product | | | 0.167* | | |
| | | | (0.100) | | |
| Hedonic Product * Modality Diversification * Engagement Diversification | | | 0.060*** | | |
| | | | (0.019) | | |
| Young Customer | | | | -0.337*** | |
| | | | | (0.060) | |
| Young Customer * Modality Diversification * Engagement Diversification | | | | 0.096*** | |
| | | | | (0.036) | |
| High-Income Customer | | | | | -0.186*** |
| | | | | | (0.025) |
| High-Income Customer * Modality Diversification * Engagement Diversification | | | | | 0.034** |
| | | | | | (0.014) |
| Other Controls | Lower-Order Interactions, Share of Facebook/Twitter/Instagram/YouTube Engagement, Platform Number, Log Total Number of Followers, Log Total Number of Postings, Log Total Number of Engagements, Company Age, Number of SKUs, Retailer FE, Year FE, Merchant Type, Product Category | | | | |
| Observations | 6,994 | 6,994 | 6,994 | 6,994 | 6,994 |
| R Squared | 0.971 | 0.971 | 0.971 | 0.975 | 0.976 |

Notes:
1. The dependent variable is the log total web sales of the focal firm in the focal year.
2. Engagement Diversification is constructed using the same method as in Tables 3 and 4. Modality Diversification is constructed using the same method as in Table 4.
3. Experience products (as opposed to search goods) are defined as products in these categories: Apparel/Accessories, Books/Music/Video, Flowers/Gifts, Food/Beverage, Health/Beauty, Jewelry, Housewares/Home Furnishings, Sporting Goods, Toys/Hobbies, Specialty/Specialty Non-Apparel. Hedonic products (as opposed to utilitarian goods) are defined as products in these categories: Apparel/Accessories, Books/Music/Video, Flowers/Gifts, Health/Beauty, Jewelry, Housewares/Home Furnishings, Toys/Hobbies.
4. Young Customer is constructed using the proportion of customers aged below 35. High-Income Customer is constructed using the proportion of customers with annual income above $100,000.
5. Lower-order interaction terms, such as Experience Product × Modality Diversification, are all added in the regression models.
6. Clustered standard errors on firm level in parentheses: *** $p<0.01$, ** $p<0.05$, * $p<0.1$.

using the Benjamini–Hochberg procedure to control the false discovery rate (Benjamini & Hochberg, 1995). All interaction effects remain statistically significant after the adjustment. We also examine potential multicollinearity among diversification measures and platform engagement shares. VIF diagnostics show that the mean VIF is 2.79 and all individual VIF values are well below the commonly used threshold of 10, indicating that multicollinearity is unlikely to affect our estimates.

### A.7 Traffic Structure Rather Than Traffic Expansion

In this section, we provide further evidence on whether the performance benefits of diversified engagement arise from expanding overall social media reach or from reshaping the structure of social-media-referred traffic. To do so, we exploit a unique dataset covering the Top 500 U.S. online retailers between 2013 and 2015, which reports detailed information on web traffic originating from major social platforms such as Facebook, Twitter, Pinterest, and YouTube to the website of the focal retailer. Unfortunately, the publisher ceased to provide such data after 2015. Using this 3-year data, we build a panel for about 500 retailers. The following table shows the summary statistics of this panel:

**Table A.7.1. Summary Statistics for Social Media Traffic Data (n=1,300)**

| Statistic | Mean | St. Dev. | Min | Max |
|---|---|---|---|---|
| Social Commerce Sale (dollars) | 8,009,656 | 35,468,483 | 4,468 | 694,346,087 |
| Facebook Traffic | 0.035 | 0.030 | 0.000 | 0.339 |
| Twitter Traffic | 0.001 | 0.003 | 0 | 0.047 |
| Pinterest Traffic | 0.003 | 0.008 | 0 | 0.110 |
| YouTube Traffic | 0.008 | 0.015 | 0 | 0.171 |
| Total Social Traffic | 0.047 | 0.036 | 0.003 | 0.351 |
| Social Traffic Diversification | 0 | 1 | -1.77 | 2.31 |

The Total Social Traffic is the traffic from all social platforms (including traffic from platforms other than Facebook, Twitter, Pinterest, and YouTube) as a percentage of the retailer's total website traffic. Social Traffic Diversification is calculated as:

$$TrafficDiv_{i,t} = 1 - \left[\left(\frac{F_{i,t}}{F_{i,t}+T_{i,t}+P_{i,t}+Y_{i,t}}\right)^2 + \left(\frac{T_{i,t}}{F_{i,t}+T_{i,t}+P_{i,t}+Y_{i,t}}\right)^2 + \left(\frac{P_{i,t}}{F_{i,t}+T_{i,t}+P_{i,t}+Y_{i,t}}\right)^2 + \left(\frac{Y_{i,t}}{F_{i,t}+T_{i,t}+P_{i,t}+Y_{i,t}}\right)^2\right]$$

where $F_{i,t}$ stands for Facebook traffic of company $i$ in year $t$, and $T_{i,t}$, $P_{i,t}$ and $Y_{i,t}$ stands for Twitter, Pinterest and Youtube, respectively.

Using these data, we construct measures of social-media-referred traffic diversification as well as sales attributable to social media traffic. The results are reported in Table A.7.2. Column (1) first examines whether engagement diversification translates into a more diversified distribution of social-media-referred traffic across platforms. The estimated coefficient is positive and statistically significant, indicating that firms allocating engagement more diversely across platforms also receive more balanced inbound traffic from those platforms. This result confirms that engagement allocation shapes the structure of social media traffic reaching the firm's website.

Columns (2) and (3) then examine the implications of this traffic structure for sales attributable to social media. Engagement diversification is positively associated with social-commerce sales, suggesting that diversified engagement strategies translate into higher revenues generated from social media referrals. Moreover, when we include social traffic diversification directly in the regression, it is also positively and significantly associated with social-commerce sales, indicating that a more diversified traffic structure itself contributes to improved performance.

In addition, Columns (4) and (5) examine the conversion rate of social-media-referred traffic. Consistent with the earlier mechanism tests in Table 6, engagement diversification is positively associated with conversion rates, suggesting that diversified engagement improves the effectiveness of converting social media visitors into customers. Finally, Columns (6) and (7) examine two alternative channels: total social media traffic and ticket size attributable to social media. In contrast to the conversion results, engagement diversification has no statistically significant effect on either the volume of social-media-referred traffic or the average transaction value, consistent with the earlier results in the main text Table 5.

Taken together, these results indicate that diversified engagement improves firm performance primarily by reshaping the composition of social media traffic and increasing the conversion efficiency of incoming visitors rather than by expanding the total volume of social media traffic. In other words, diversification enhances the allocation efficiency of social media attention rather than merely increasing overall exposure.

**Table A.7.2. Mechanism Test: Diversification Improves Sales and Conversion by Changing Traffic Structure Rather Than Expanding Total Traffic**

| Model | OLS (1) | OLS (2) | OLS (3) | OLS (4) | OLS (5) | OLS (6) | OLS (7) |
|---|---|---|---|---|---|---|---|
| Dependent Variable | Social traffic diversification | Log web sales attributable to social media traffic | | Log conversion rate attributable to social media traffic | | Log total social media traffic | Log ticket size attributable to social media traffic |
| | | | | | | | |
| Engagement Diversification | 0.592*** | 1.091*** | 0.0061 | 0.0030*** | 0.0009 | -0.038 | -0.018 |
| | (0.016) | (0.029) | (0.0108) | (0.0005) | (0.0009) | (0.053) | (0.012) |
| Social traffic diversification | | | 1.8318*** | | 0.0034*** | | |
| | | | (0.0148) | | (0.0012) | | |
| Share of Facebook Engagement | -0.009** | -0.016* | 0.0005 | -0.0001 | -0.0001 | -0.029** | -0.001 |
| | (0.004) | (0.008) | (0.0018) | (0.0001) | (0.0001) | (0.014) | (0.003) |
| Share of Twitter Engagement | 0.003 | 0.006 | -0.0000 | 0.0001 | 0.0000 | 0.010 | 0.001 |
| | (0.003) | (0.006) | (0.0012) | (0.0001) | (0.0001) | (0.010) | (0.002) |
| Share of Pinterest Engagement | 0.003 | 0.008 | 0.0023** | -0.0001 | -0.0001* | 0.007 | 0.003 |
| | (0.003) | (0.005) | (0.0011) | (0.0001) | (0.0001) | (0.009) | (0.002) |
| Other Controls | Platform Number, Log Total Number of Followers, Log Total Number of Postings, Log Total Number of Engagements, Company Age, Number of SKUs, Retailer FE, Year FE, Merchant Type, Product Category | | | | | | |
| Observations | 1,300 | 1,300 | 1,300 | 1,300 | 1,300 | 1,300 | 1,300 |
| R Squared | 0.987 | 0.988 | 0.999 | 0.965 | 0.965 | 0.955 | 0.986 |

Notes:

1. The dependent variable in Column (1) is the diversification of social-media-referred web traffic, measured as one minus the Herfindahl–Hirschman Index (HHI) of the distribution of traffic across social media platforms. The dependent variable in Columns (2) and (3) is the log total web sales attributable to social media traffic. The dependent variable in Columns (4) and (5) is the log conversion rate (i.e., the percentage of users who make a transaction out of the total number of visitors) attributable to social media traffic. The dependent variable in Column (6) is the log of total traffic (i.e., the number of visits to the firm’s website) attributable to social media. The dependent variable in Column (7) is the log of ticket size (i.e., the average dollar amount users spend per transaction) attributable to social media traffic. By definition, log web sales = log total traffic + log ticket size + log conversion rate.
2. Engagement Diversification is constructed using the same method as in Tables 3 to 6. Social traffic diversification measures the diversification of social-media-referred web traffic, measured as one minus the HHI of the distribution of traffic across social media platforms. Share of Facebook/Twitter/Instagram/YouTube Engagement is constructed using the same method as in Tables 3 to 6.
3. All columns report the OLS results with platform number, log total number of followers, log total number of postings, log total number of engagements, company age, number of SKUs, company fixed effects, and year fixed effects.
4. Clustered standard errors on firm level in parentheses: *** $p<0.01$, ** $p<0.05$, * $p<0.1$.

## A.8 Traffic Quality Evidence

Table A.8.1 evaluates traffic quality metrics. Starting in 2018, Internet Retailer Guide began publishing website traffic analytics, including bounce rates, visit duration, and pages per visit. We use these indicators as dependent variables in Table A.8.1. The results show that diversification reduces bounce rates, increases visit duration, and increases pages per visit. The interaction between engagement diversification and modality diversification further strengthens these effects. Taken together, these results demonstrate that diversification enhances traffic quality, consistent with improved cognitive processing and decision facilitation.

**Table A.8.1. Mechanism Test: Diversification Improves Traffic Quality**

| Model | OLS (1) | OLS (2) | OLS (3) | OLS (4) | OLS (5) | OLS (6) |
|---|---|---|---|---|---|---|
| Dependent Variable | Bounce Rate | | Visit Duration | | Pages per Visit | |
| | | | | | | |
| Engagement Diversification | -0.0020*** | -0.0010 | 2.754*** | 0.899* | 0.041*** | 0.002 |
| | (0.0007) | (0.0009) | (0.370) | (0.470) | (0.007) | (0.009) |
| Modality Diversification | | -0.0001 | | 1.878*** | | 0.038*** |
| | | (0.0001) | | (0.480) | | (0.009) |
| Modality Diversification * Engagement Diversification | | -0.0013** | | 1.094*** | | 0.025*** |
| | | (0.0006) | | (0.319) | | (0.006) |
| Other Controls | Share of Engagement on Facebook, Twitter, Instagram, and YouTube, Platform Number, Log Total Number of Followers, Log Total Number of Postings, Log Total Number of Engagements, Company Age, Number of SKUs | | | | | |
| Observations | 1,949 | 1,949 | 1,949 | 1,949 | 1,949 | 1,949 |
| R Squared | 0.036 | 0.038 | 0.104 | 0.123 | 0.128 | 0.150 |

Notes:
1. The dependent variable in Columns (1) and (2) is the bounce rate (the percentage of visitors who enter the website and leave after viewing only one page, without interacting with the site) of the web traffic of the focal company in the focal year. The dependent variable in Columns (3) and (4) is the average visit duration (the total time, in seconds, a visitor spends on the website) for the focal company's web traffic in the focal year. The dependent variables in Columns (5) and (6) are the average pages per visit (the total number of web pages a visitor browses on the focal company's website) for the focal year's web traffic.
2. Engagement Diversification is constructed using the same method as in Tables 3 to 6. Modality Diversification is constructed using the same method as in Tables 4 to 6. Share of Facebook/Twitter/Instagram/YouTube Engagement is constructed using the same method as in Tables 3 to 6.
3. All columns report the OLS results without fixed effects because only 2 years of data are available for the metrics of bounce rate, visit duration, and pages per visit.
4. Clustered standard errors on firm level in parentheses: *** $p<0.01$, ** $p<0.05$, * $p<0.1$.

## A.9 Distribution of Overlapping Index and The Marginal Effect of Engagement Diversification Conditioning on Overlapping Index

The following figures show the distribution of the overlapping index variable (after mean-centering) and the marginal effect of engagement diversification conditioning on the overlapping index..

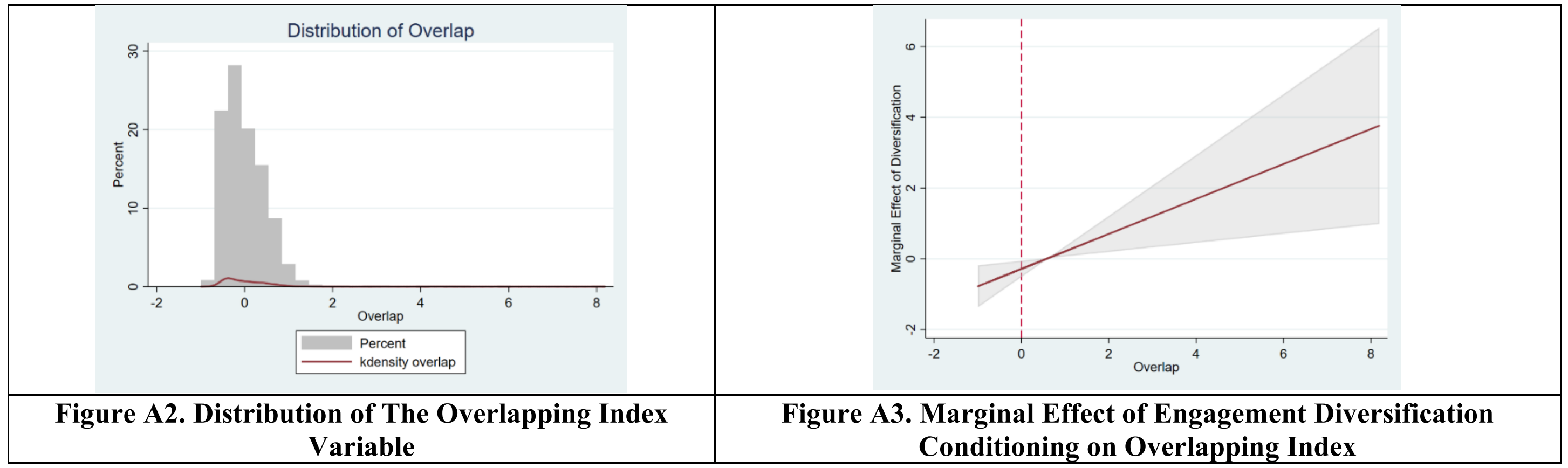

| Figure A2. Distribution of The Overlapping Index Variable | Figure A3. Marginal Effect of Engagement Diversification Conditioning on Overlapping Index |
|---|---|

In the following table, we also show that the memory reinforcement effect results remain directionally consistent using the winsorized (at the 5th and 95th percentiles) sample.

**Table A.9.1 Memory Reinforcement Effect Using Winsorized (at the 5th and 95th percentiles) Sample**

| Model | OLS | OLS | OLS | Policy IV 2SLS | Hausman IV 2SLS | OLS | OLS | OLS | Policy IV 2SLS | Hausman IV 2SLS |
|---|---|---|---|---|---|---|---|---|---|---|
| | (1) | (2) | (3) | (4) | (5) | (6) | (7) | (8) | (9) | (10) |
| Dependent Variable | Log web sales | | | | | Log conversion rate | | | | |
| | | | | | | | | | | |
| Engagement Diversification (On Twitter and Pinterest) | 0.054*** | -0.009 | -0.003 | **-0.105*****  | **0.023** | 0.0007*** | -0.0001 | -0.0001 | **-0.0012*****  | **0.0002** |
| | (0.010) | (0.013) | (0.013) | **(0.029)** | **(0.047)** | (0.0001) | (0.0002) | (0.0002) | **(0.0003)** | **(0.0006)** |
| Overlapping | | 0.216*** | 0.225*** | **0.076****  | **0.262*****  | | 0.0027*** | 0.0028*** | **0.0009** | **0.0032*****  |
| | | (0.028) | (0.028) | **(0.038)** | **(0.072)** | | (0.0004) | (0.0004) | **(0.0006)** | **(0.0010)** |
| Engagement Diversification * Overlapping | | | 0.068*** | **0.649*****  | **0.213*****  | | | 0.0005*** | **0.0079*****  | **0.0023*****  |
| | | | (0.020) | **(0.169)** | **(0.070)** | | | (0.0002) | **(0.0021)** | **(0.0004)** |
| Average Visits Per User | 0.017*** | 0.017*** | 0.018*** | 0.018*** | 0.017*** | 0.0020*** | 0.0021*** | 0.0020*** | 0.0020*** | 0.0021*** |
| | (0.003) | (0.003) | (0.003) | (0.003) | (0.003) | (0.0005) | (0.0005) | (0.0006) | (0.0005) | (0.0004) |
| Other Controls | Log Number of Twitter Engagement, Log Number of Pinterest Engagement, Platform Number, Log Total Number of Followers, Log Total Number of Postings, Company Age, Number of SKUs, Retailer FE, Year FE, State-Year FE, Merchant Type, Product Category, Product Category-Year FE | | | | | | | | | |
| Observations | 6,191 | 6,191 | 6,191 | 6,191 | 6,191 | 6,191 | 6,191 | 6,191 | 6,191 | 6,191 |
| R Squared | 0.969 | 0.970 | 0.970 | \ | \ | 0.969 | 0.970 | 0.970 | \ | \ |

Notes:
1. The dependent variable in Columns (1) – (5) is the log total web sales of the focal firm in the focal year. The dependent variable in Columns (6) – (10) is the log conversion rate of the focal firm in the focal year. The sample is winsorized at the 5th and 95th percentiles.
2. Engagement Diversification is constructed as one minus the HHI of the distribution of likes received across the 2 social platforms (Twitter and Pinterest) each year. To ensure comparability across platforms, we standardize the distribution of likes on each platform by dividing it by the platform-year median before computing the HHI. This adjustment reduces distortions arising from differences in platform size and engagement intensity.
3. Overlapping is constructed as $OverlappingIndex_{i,t} = \frac{\#SamePerson_{i,t}}{\#TwitterFollower_{i,t} + \#PinterestFollower_{i,t}}$, where for each company and year, we obtain the full follower lists on Twitter and Pinterest, and treat 2 follower IDs as the same person if their Jaro-Winkler distance is below 10%. Then we calculate the proportion of the same person over the total number of followers on both platforms as the overlapping index.
4. Average visits per user is constructed as the ratio of the total number of web visits to the number of unique visitors.
5. Columns (1) – (3) and (6) – (8) report the OLS results with company and year fixed effects. Columns (4) and (9) report the 2SLS results using platform policies as instrumental variables. Columns (5) and (10) report the 2SLS results using Hausman-type instruments as instrumental variables. The first-stage F-statistic is greater than 20. Variables in bold are treated as endogenous and estimated using IV 2SLS.
6. Clustered standard errors on firm level in parentheses: *** p<0.01, ** p<0.05, * p<0.1.